\documentclass[aps,prd, showpacs, amsmath,amssymb,eqsecnum]{revtex4}
\usepackage{graphicx}%
\usepackage{subfigure}
\usepackage{mathptmx}
\usepackage{bm}
\newcommand{\beq}{\begin{equation}}
\newcommand{\eeq}{\end{equation}}

\newcommand{\be}{\begin{eqnarray}}
\newcommand{\ee}{\end{eqnarray}}
\long\def\hidestart#1\hideend{}
\begin{document}

\title{On Scale Determination in Lattice QCD with Dynamical Quarks}

\author{Asit K. De}
\email{asitk.de@saha.ac.in}

\author{A. Harindranath}
\email{a.harindranath@saha.ac.in}

\author{Jyotirmoy Maiti}
\email{jyotirmoy.maiti@saha.ac.in}

\affiliation{Theory Group, Saha Institute of Nuclear Physics \\
 1/AF Bidhan Nagar, Kolkata 700064, India}

\date{March 8, 2008}

\begin{abstract}
Dependence of $a/r_c$ (inverse Sommer parameter in units of lattice spacing $a$) on $am_q$ (quark mass in lattice unit) has been observed in all lattice QCD simulations with sea quarks including the ones with improved actions. How much of this dependence is a scaling violation has remained an intriguing question. Our approach has been to investigate the issue with an action with known lattice artifacts, i.e., the standard Wilson quark and gauge action with $\beta=5.6$ and 2 degenerate flavors of sea quarks on 
$ 16^3 \times 32 $ lattices. In order to study in detail the
sea quark mass dependence, measurements are carried out at eight values
of the Wilson hopping parameter $\kappa$ in the range 0.156 - 0.158 corresponding to PCAC quark mass values $am_q$ from about 0.07 to below 0.015.
We analyze the static potential by fitting to the familiar phenomenological
form and extract $a/r_c$.
Though scaling violations may indeed be present for
relatively large $am_q$, a consistent scenario at sufficiently small $am_q$ seems to emerge in the mass-independent scheme where for a fixed $\beta$, $1/r_0$ and
$\sqrt{\sigma}$ have linear dependence on $m_q$ as physical effects similar to the quark mass dependence of the rho mass.
We present evidence for this scenario and accordingly
extract the lattice scale $a$ by chiral extrapolation to the physical point. Care has been exercised to determine optimal values of all fitting parameters and accuracy of the chiral extrapolation. An independent determination of the scale $a$ by chiral extrapolation of the rho mass is consistent with the scale obtained above ($a$ = 0.08041(12)(77) fm, 
$a^{-1}$ = 2.454(4)(23) GeV).

\end{abstract}

\pacs{02.70.Uu, 11.10.Gh, 11.10.Kk, 11.15.Ha}

\maketitle


\section{Introduction}\label{intro}
An accurate determination of the lattice scale is mandatory for
comparing lattice observables with their continuum counterparts. 
While the determination of lattice scale is
conceptually simple in the quenched approximation of lattice QCD, simulations with 
dynamical quarks bring forth some unavoidable complications. It is of interest to study whether
some of the complications are due to lattice artifacts  
or they throw light on physical aspects of lattice QCD.    

In the early days of lattice QCD, string tension provided a method
to set the scale. 
If the  color electric flux emanating from the quarks are squeezed into 
one dimension (a string-like configuaration),
such a gauge field configuartion will have constant energy
per unit length i.e., $E = \sigma r$. The energy density $\sigma$ is
called the string tension. For the quenched theory, string tension is
a fundamental quantity. For a review, see, Ref.   \cite{marinari}.    
In lattice gauge theory, string tension is defined as the asymptotic
value of the force $F(r)$ between a pair of static quark and 
anti-quark \cite{balirev} at separation $r$, i.e.,
string tension $\sigma = {\rm limit}_{r \leftarrow \infty} F(r)$. The limiting
value may not be easy to extract from the lattice data since
statistical errors on the force (which is extracted from the
large Euclidean time  exponential fall-off of the expectation value of
the Wilson loop) increases with the distance. 

In the presence of dynamical quarks, the string may break at large $r$ and the definition
of string tension as the asymptotic value of the force between static
sources may not be useful. However, if we {\em define} the string
tension to be the coefficient of the linearly rising potential (which
in principle exists at all quark antiquark separations within the
hadron i.e., for less than a fermi (fm)), then this notion can survive the
presence of dynamical quarks and may be used for practical purposes.  
If we use the string tension to set the scale, we need to know the
value of the string tension in physical dimensions and there exists a fair degree of uncertainty about this value 
\cite{balirev,ape,bs932,sesam96,cppacs98,milc1}. In addition, it was also noticed with the introduction of sea quarks that the string tension in lattice units has a dependence on the quark mass in lattice units.

In Ref. \cite{sommer} Sommer introduced  the method of determining the lattice spacing
through distance scales $r_c$ derived from the potential between a static quark-antiquark pair, e.g. ,
using phenomenologically reasonably well-known information of $r_0^2 F(r_0)$=1.65 where $r_0=0.49$ fm.  Also see Ref. \cite{guagnelli}.
The Sommer scale $r_0$ was originally conceived as a
{\em bosonic observable} and was expected to be independent of the sea quarks. 
However, all lattice QCD simulations with dynamical quarks employing a variety of actions (including improved quark and gauge actions) have shown that the ratio $r_0/a$, i.e., the Sommer scale in units of the lattice spacing $a$ depends significantly on $am_q$, the quark mass in lattice units.

In this work we carry out a detailed and careful investigation of the
static potential and the ratio $a/r_c$, where $r_c$ is a Sommer-type
scale, in a lattice QCD simulation with standard Wilson gauge and
quark actions on $16^3 \times 32$ lattices at a single gauge coupling parameter $\beta = 6/g^2\,=\,5.6$ with a large set (eight values) of the fermionic hopping parameter  $\kappa$ for the sea quarks. Use of this large set of sea quark masses has helped us to identify, in terms of a certain parameterization of the static potential, the scale-violating part of the dependence of $a/r_c$ on $am_q$. Our numerical data support the interpretation that for small enough $am_q~(\lesssim 0.035)$ the dependence of $a/r_c$ on $am_q$ is a physical effect. 

In a mass-independent scheme, something that follows quite naturally as explained later in this paper, the scale $a$ is then obtained by chirally extrapolating $a/r_c$ to the physical point. Since no other information is available on the dependence of $a/r_c$ on $am_q$ other than from numerical simulations, this extrapolation needs to be done with care to exclude large uncertainties from the extrapolation. Ignoring possible uncertainties of the values of $r_c^{\rm ph}$ ($r_c$ at the physical point), we find accurate values of the scale determined this way with about $1\%$ error.  

Independent of the determination from the static potential and the Sommer scale, we have also determined the scale $a$ from chiral extrapolation of $am_\rho$, the rho meson mass in lattice unit. The scale determined this way is consistent with the scale determined from the chiral extrapolation of $a/r_c$, although with somewhat larger error bars ($\sim 2-2.5\%$).

In our determination of the static potential and the subsequent analysis to obtain the ratio $a/r_c$, we have exercised utmost care in determining the fit range to determine the static potential, the fit range to determine the parameters of the static potential and also the optimum smearing levels to be used for the gauge configurations. In addition, for the chiral extrapolations of $a\sigma^{1/2}$ and $a/r_c$, we have first used $am_q$ and then $(am_\pi)^2$ as the chiral regulator to double check the reliability of the extrapolation. We have preferred $(am_\pi)^2$ to $(r_cm_\pi)^2$ for chiral extrapolation to the physical point because $r_c$ itself has a chiral dependence. We have used all possible cross-checking of different determinations for consistency of our results and we present evidence for consistency in this paper.

Our results are based on accurate determinations of the parameters of the static potential and reliable chiral extrapolations. We believe that our qualitative conclusions, if not also the quantitative conclusions, are independent of the numerical details like the fit-ranges, smearing levels etc.      

In an earlier paper \cite{paper0}, at the same set of parameters with the same action we have determined, using gaussian smearing both on source and sink, the pion and the rho masses, their decay constants and the PCAC quark mass. We have used most of those results in this paper. For details, please see Ref. \cite{paper0}.

Before we end this section, we would like to point out that there exists in the literature a variety of other methods to determine the lattice scale 
\cite{sesam,domain,deldebbio1,davies97,luescher2006}.

This paper is organized as follows. Section \ref{simu} contains a
summary of the simulation.  In Sec. \ref{wloop} measurement of
the Wilson loop and the extraction of the static potential are given and 
Sec. \ref{fit} describes the fit of the static potential. Sommer 
scale is described in Sec. \ref{sommer}. Dependence of various
parameters of the potential on the lattice quark mass $am_q$ is
detailed in Sec. \ref{depamq} and the interpretation of the $am_q$
dependence is presented in Sec. \ref{interpret}. Section \ref{chiral}
presents the extrapolation of the data to chiral and physical points
and the determination of the lattice scale at the physical point. 
Sec. \ref{sigmad} presents estimates of the physical string
tension. Implications from weak coupling perturbation theory are
explored in Sec. \ref{impli}. Finally, Sec. \ref{concl} discusses the 
salient features and the conclusions.

\section{Simulation}\label{simu}
We have used {\em unimproved Wilson gauge and
fermion actions} at a single gauge coupling given by $\beta=6/g^2\,=\,5.6$ and two flavors of degenerate sea quarks on $16^3 \times 32$ lattices. Our
choice of the gauge coupling is motivated by the
requirements of a reasonably small lattice spacing so that the results of the simulation are not significantly contaminated by scaling vilolations. 
As mentioned already in the Introduction, lattice QCD simulations in the past using various gauge and fermion actions have observed significant sea quark mass dependence on $a\sigma^{1/2}$ and $a/r_c$.
In order to study this sea quark mass dependence
in detail, we have generated gauge configurations and performed measurements 
at a large set of sea quark masses corresponding to eight values of the
Wilson hopping parameter $\kappa$ = 0.156, 0.1565, 0.15675, 
0.157, 0.15725, 0.1575, 0.15775 and 0.158. 

At each $\kappa$ we have generated 5000 equilibrated configurations 
with the standard HMC algorithm (with even-odd pre-conditioned
Conjugate Gradient for inversion of $M^{\dagger}M$, $M$ being the fermion matrix)
and performed Wilson loop measurements separated by 
25 configurations. Details on the simulation  and autocorrelation times
can be found in Ref. \cite{paper0}. 

In Ref. \cite{paper0} we have presented a detailed study of the pion and the rho mass, their decay constants and PCAC quark mass using the same set of simulation parameters. We used gaussian smearing on both source and sink and investigated systematic effects on the pion mass and the decay constant using different types of correlators (PP, AA, AP and PA, where P and A respectively are pseudoscalar and axial vector densities). PCAC quark masses were determined using different pion correlators. We observed finite size (FS) effects on our lowest pion masses. However, interestingly the effect was different for different operators, e.g., pion mass from the AA correlator at $\kappa=0.15775$ had negligible FS effect  while the same computed from the PP correlator had significant FS effect. At $\kappa=0.158$ (our largest value of $\kappa$), pion masses computed from either operator had significant FS effect, but the effect was less in the AA correlator.

In this paper, we shall study the $am_q$ dependence of $a/r_c$ and other quantities. Eventually the pion masses in lattice units are used for chiral extrapolation of quantities derived from the static potential to the physical point. We use all results of quark and pion mass obtained in Ref. \cite{paper0} in this paper. We use similar notation and convention as developed in \cite{paper0}, e.g., lattice quark masses are denoted $am_q^{AA}$ or $am_q^{AP}$ depending on the correlator used and each of these quark masses further depend on whether the pion mass used in the determination of the quark mass was taken from the PP or the AA correlator.

Unless otherwise stated, all errors in this paper presented in data (in text or tables) or shown in figures are single-omission jackknife errors computed from 200 jackknife bins.
\section{Extraction of the Static 
Potential from Wilson loops}\label{wloop}
For sufficiently large times $T$, the asymptotic behavior of the 
expectation value of the Wilson loop $W(R,T)$ ($R$ being the spatial separation) 
is given by $
\langle W(R,T)\rangle~=~ C(R) ~{\rm exp}\big [{-V(R)T} \big ]~$, where
$V(R)$ is the potential between a pair of static quark and antiquark. The 
coefficient $C(R)$ is the ground state overlap.
In order to determine the static potential reliably, it is
important to have a large ground state overlap in the measurement of
the Wilson loop.

The Wilson loop is a gauge-invariant quantity and as such gauge-fixing
the gauge configurations is {\em not} necessary for its measurement,
although we do it anyway.
After gauge fixing to temporal gauge \cite{creutz}, APE smearing \cite{ape}
is performed on the link fields.
Smearing gets rid of short distance fluctuations, helps in
reducing higher states' contamination and increase overlap
with physical states of interest. For a very early discussion of the
need for smearing, see Ref. \cite{parisi}. Smearing of gauge fields in a
fermion action is also expected to have several other advantages \cite{cbtd}.

APE smearing is performed as follows:
\begin{eqnarray}
U_i(x)\rightarrow U^{\prime}_i(x) = 
 (1 - c) U_i(x) + 
\frac{c}{4}
\sum_{\rm staples}\tilde{U}_i(x),
~~~{\rm where}~{\tilde{U}_i(x)}  {=} {
  U_j(x+i)U^{\dagger}_i(x+j)U^{\dagger}_j(x)},
\end{eqnarray}
followed by projection back to $SU(3)$. 
The parameter $c$ is the relative strength of the smearing and 
we have chosen $c=4/(\epsilon
+4)$ with $\epsilon$=2.5.

We measured the Wilson loops $\langle W(R,T)\rangle$ with temporal
extents up to $T=16$ and
spatial separations up to $R=\sqrt{3} \times 8$.

A reasonable estimate of the static potential $aV(R)$ is obtained by
the plateau reached at large $T$ of the {\em effective} potential
\begin{equation}
aV_{\rm eff}(R,T) = 
{\rm ln}\frac{\langle W(R,T)\rangle}{\langle W(R,T+1)\rangle}. 
\end{equation} 

\begin{figure}
\begin{center}
\includegraphics[width=.8\textwidth]{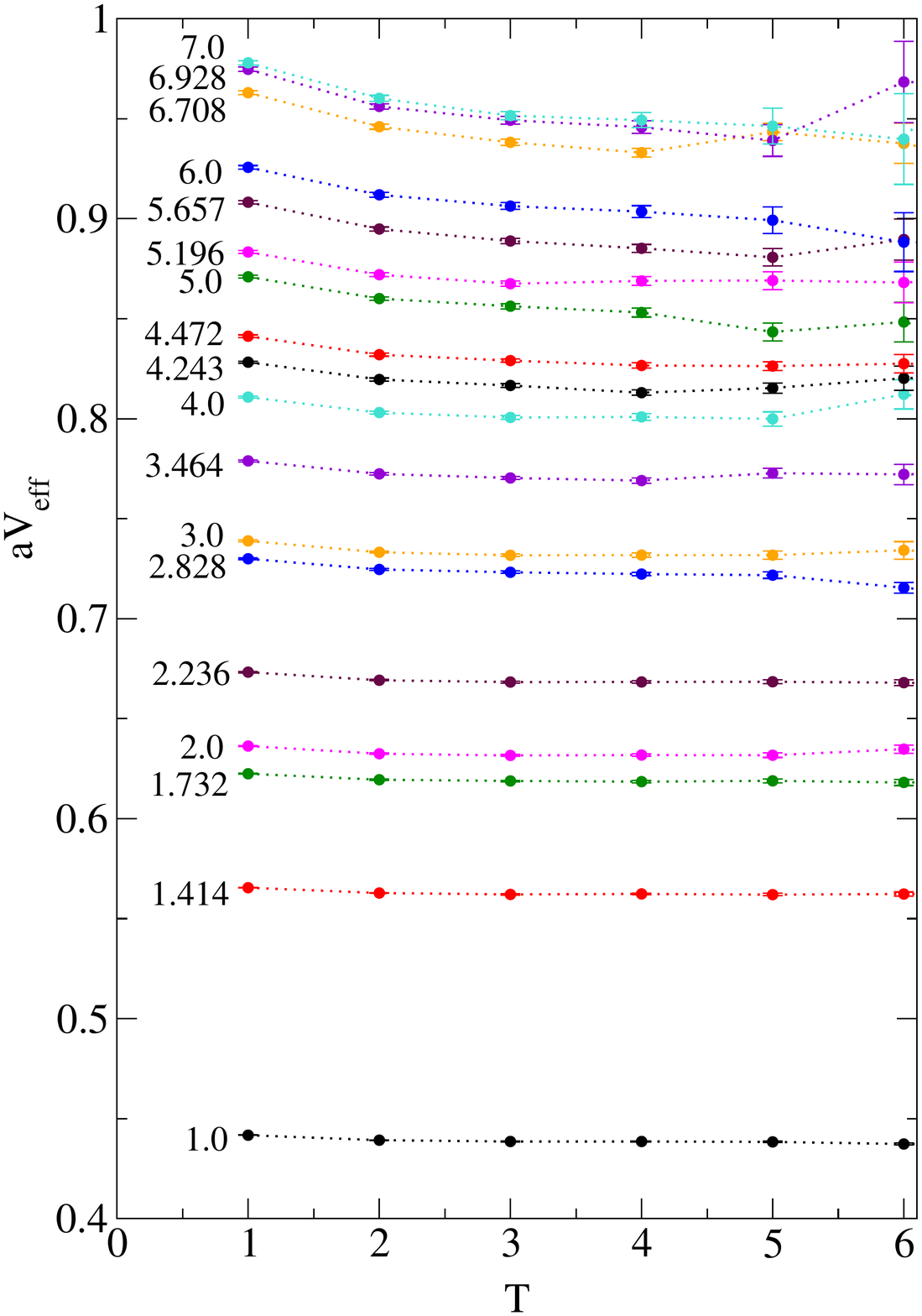}
\end{center}
\caption{The {\em effective} static potential $aV_{\rm eff}(R,T)$ as
a function of $T$ for different values of 
$R$ at smear level 30 and $\kappa$=0.15775. }
\label{eff_pot}
\end{figure}

Fig. \ref{eff_pot} shows, at $\kappa=0.15775$ and smearing level 30,
$aV_{\rm eff}$ as a function of $T$ for a host of values of $R$
ranging from $R=1$ to $R=7$. Generally plateaux are observed in this
figure starting from $R=3$. As $T$ grows for a given $R$, or as $R$
grows, the data get noisy. 

\begin{figure}
\begin{center}
\includegraphics[width=.8\textwidth]{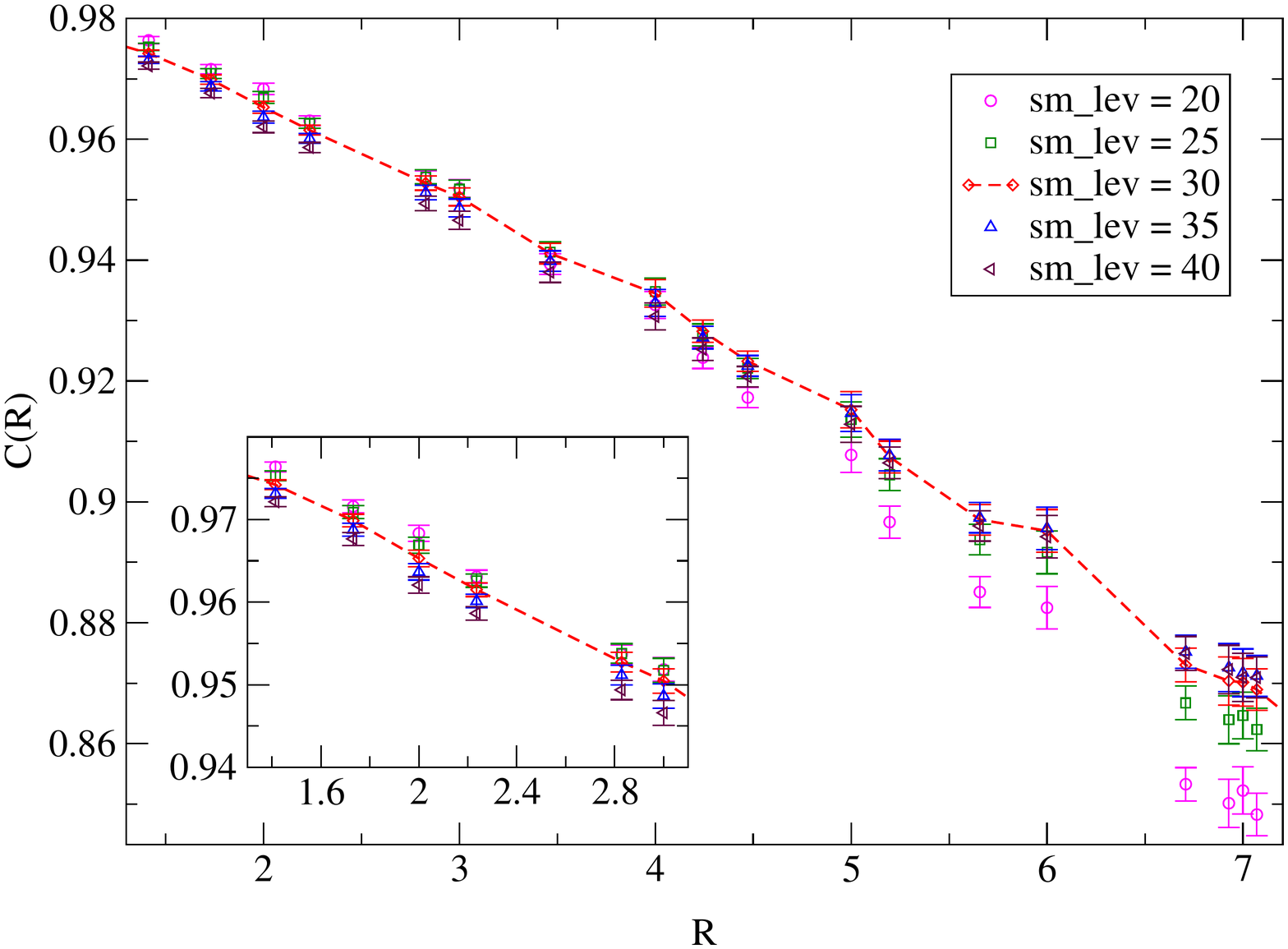}
\end{center}
\caption{Ground state overlap $C(R)$ versus $R$ for $\kappa$=0.15775
  as a function of the smearing level (sm\_lev). The inset shows the enlarged view
  of the small $R$ region. The dashed line connecting the data points at sm\_lev = 30 shows the optimum nature of this choice both at large and small $R$ region.}
\label{CR}
\end{figure}

At each $\kappa$ the optimum level of smearing (which in the case of
$\kappa= 0.15775$ is 30 for the data shown in Fig. \ref{eff_pot}) is
obtained by comparing the ground state overlap $C(R)$ for different
smearing levels. This is what is shown in Fig. \ref{CR}. At large
values of $R$, the ground state overlap increases as the smearing
level increases (upto a certain smearing level), but at small
$R$ (as shown in the inset) the reverse is generally true. In order to have the
optimum ground state overlap we have chosen 30 as the smearing level
at this $\kappa$. For accurate analysis of $aV(R)$ in a given range of
$R$, in terms of
parameters sensitive to either small $R$ or large $R$, it is important
to choose the smearing level which gives optimum ground state overlap
throughout the range of $R$ used in the analysis of $aV(R)$.   

The optimum smearing levels used in this paper are 25 (for
$\kappa~=~0.156$, 0.1565, 0.15675, 0.157, 0.15725) and 30 (for
$\kappa~=~0.1575$, 0.15775, 0.158).

For each value of $R$, we determine $V(R)$ by a single exponential fit 
in the $T$ range $[T_{\rm min},T_{\rm max}]~=~[3,4],~ [3,5]$ and
$[4,5]$. The single exponential fitting ansatz and the fitting ranges are well justified by the plateaux in
Fig. \ref{eff_pot}. Eventually we have chosen the range $[3,4]$ for
the final analysis, as discussed later.

\section{Fit of the Static Potential}\label{fit}
\begin{figure}
\begin{center}
\includegraphics[width=.8\textwidth]{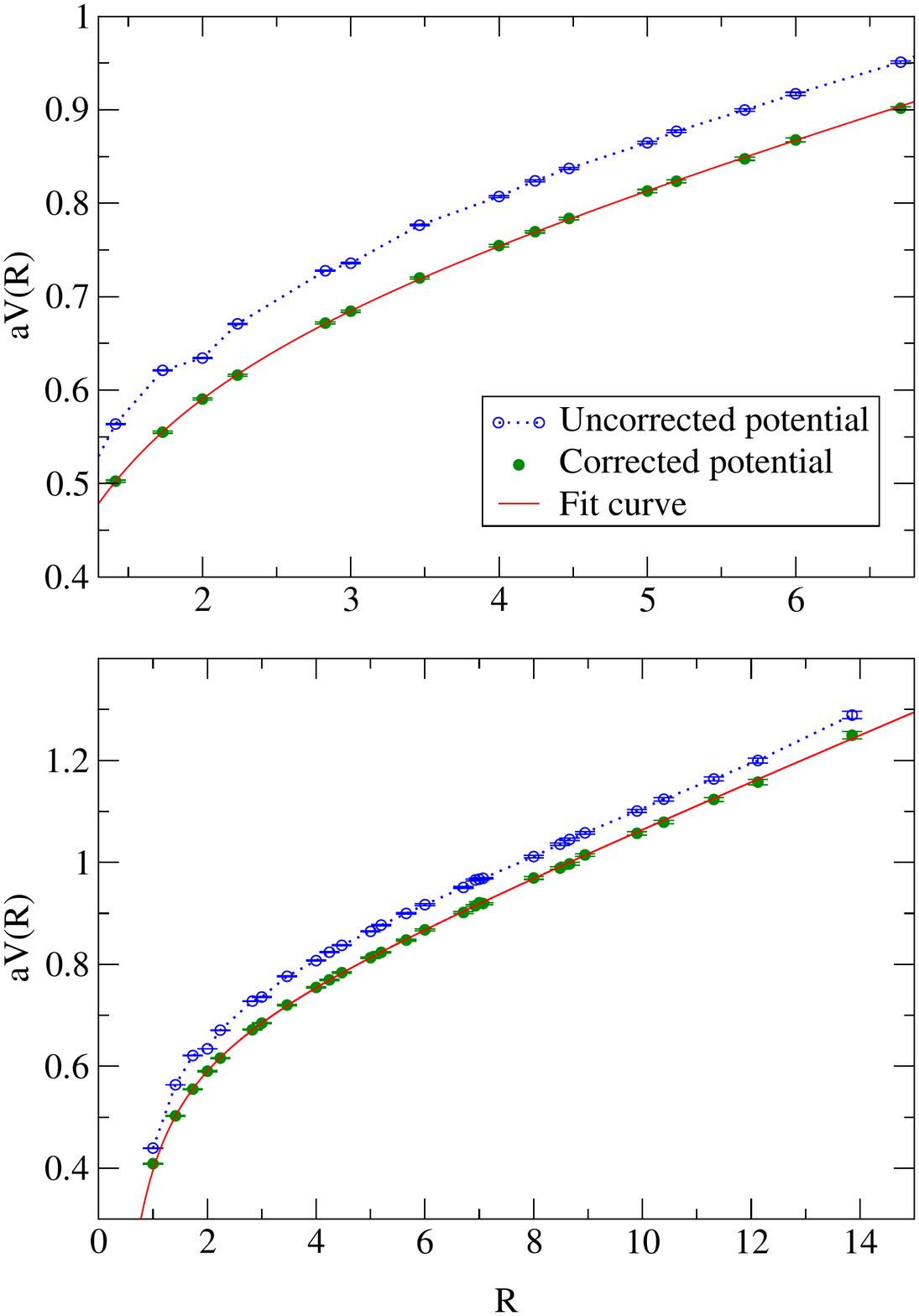}
\end{center}
\caption{Potental with and without the correction for finite lattice 
for $\kappa$= 0.1575. The upper panel shows the fitted range while the
lower panel further shows the quality of the fit in regions of $R$
beyond the fitted range.}
\label{pot-corr-1575}
\end{figure}

\begin{figure}
\begin{center}
\includegraphics[width=.8\textwidth]
{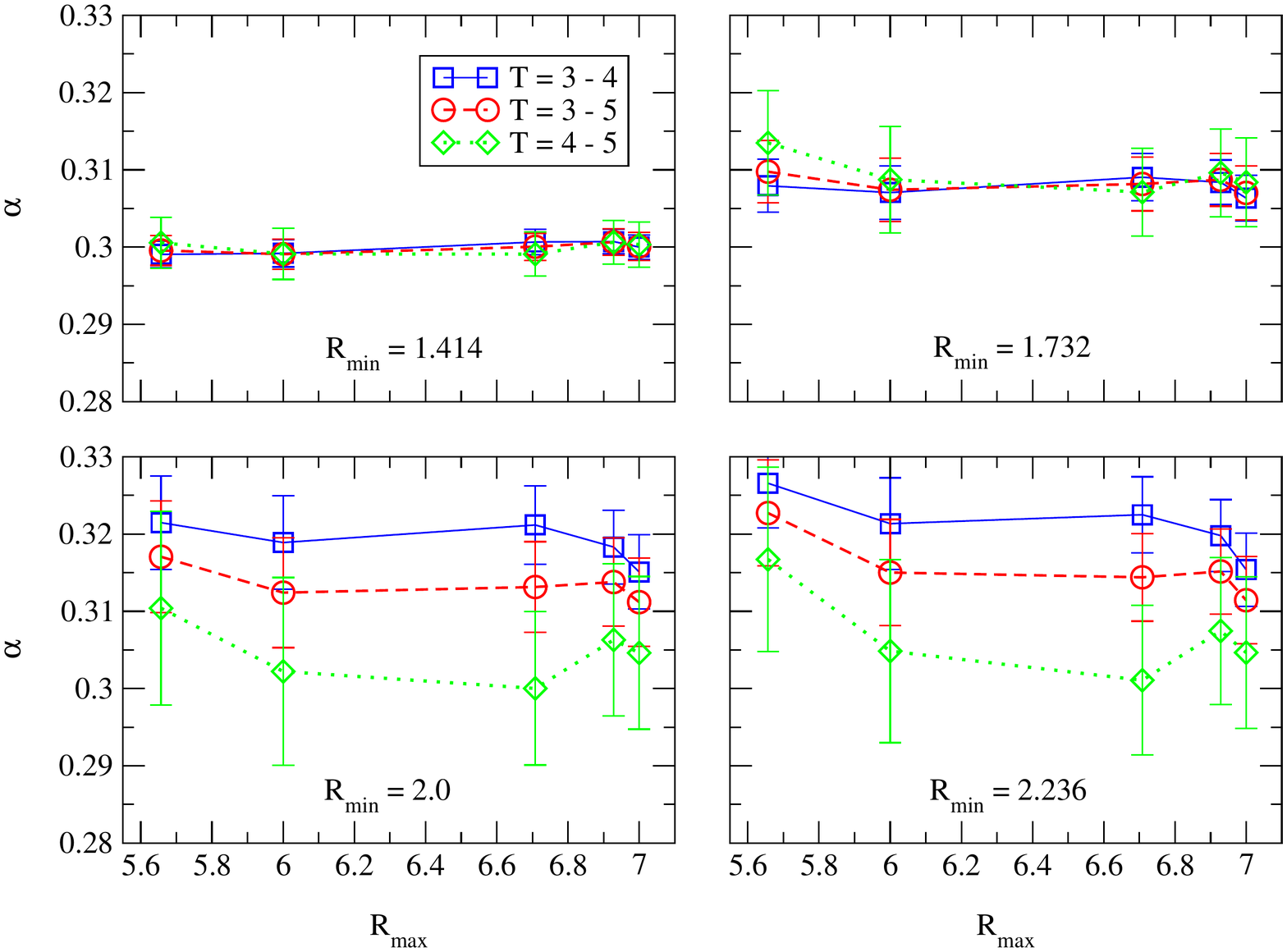}
\end{center}
\caption{The parameter $\alpha$ for $\kappa=$0.1575 as a function of 
$R_{\rm max}$ for different choices of $R_{\rm min}$ for three
  different data sets corresponding to three $T$ ranges.}
\label{alpharmaxrmin}
\end{figure}
Phenomenologically \cite{eichten}, the potential $V$ between
a static quark and antiquark is parameterized as follows:
$V(r)\, =\, V_0\, + \,\sigma\,r\, +\, \frac{\alpha}{r} $
where $\sigma$ is the string tension which has the dimension of
mass$^2$.  In lattice units, we have
$a V(r) \,=\, a V_0 \,+\, a^2 \sigma\,\frac{r}{a} \,+\, {\alpha}\frac{a}{r}
$. Writing $ r=Ra$ and $\sigma = {\tilde \sigma}/a^2$, we get
$a V(R) \,=\, a V_0 \,+\, {\tilde \sigma} R\,+\, \frac{\alpha}{R}$.

After incorporating the correction for the finite lattice using the 
expression for the 
perturbative lattice Coulomb potential \cite{langrebbi, michael}
\be
\left[\frac{1}{R}\right]~=~ \frac{4 \pi}{L^3}~ \sum_{q_{i}\neq 0}~ 
\frac{{\rm cos} (a q_{i}\cdot R)}{4 {\rm sin}^2 (a q_{i}/2)},
\ee
the parameterization of the potential on the lattice reads
\be  aV(R) ~=~ aV_0 ~+~ {\tilde \sigma}~R 
~-~ \frac{\alpha}{R}~ - ~ \delta_{\rm ROT}~ 
\left ( \Bigg[\frac{1}{R}\Bigg] - \frac{1}{R}\right )\label{corr-pot}
\ee
where $\delta_{\rm ROT}$ is the coefficient of the correction term.
The measured static potential is fit to the formula in Eq. (\ref{corr-pot})
which corrects the lattice data for the lattice artifacts in the Coulomb 
potential. The first three terms of Eq. (\ref{corr-pot}) now
gives the continuum potential (i.e., without lattice artifacts).

There are a few comments on the $1/R$ terms proportional to $\alpha$ and
$\delta_{\rm ROT}$: (i) the lattice version of $1/R$ emerges out of
fourier transforming the gluon propagator $1/q^2$ in a finite box for 1-gluon exchange interaction between a pair of static quark and antiquark and
as such the parameter $\alpha$ has the interpretation of being
proportional to the strong coupling constant, (ii) the difference 
$( [1/R] - 1/R )$ between the lattice version and
the continuum version is never negligible on a finite lattice. As a
result the correction is never very small as evident in 
Fig. \ref{pot-corr-1575}, (iii) the parameter $\alpha$ is expected to
run with $R$ at these intermediate length scales, (iv) we can then
only estimate an average $\alpha$ over
the values of $R$ where the static potetial is fit, (v) perturbative running
is generally applicable at scales $\gtrsim 2$ GeV which translates into
$R\lesssim 1$ in our case. 

Fig. \ref{pot-corr-1575} shows in 2 plots at $\kappa=0.1575$ 
the uncorrected potential as
obtained from our numerical simulation and the corrected (continuum)
potential obtained by subtracting out the correction 
($\delta_{\rm ROT}$) term. The open and solid symbols are respectively
the uncorrected and the corrected potential. The dotted straight lines
are just joining the uncorrected points. The solid lines going through
the corrected data represent the fit and by definition it should go
through the corrected points within the $R$-range of the fit which in
this case is
from $R_{\rm min}=\sqrt{2}\simeq 1.4$ to $R_{\rm max}=3\sqrt{5}\simeq
6.7$. The purpose of the upper plot is to show that the kinks at small
$R$ region due to breaking of rotational invariance in our finite
lattice disappear after the correction. The lower plot shows that even
much beyond the fit-range the fit to the corrected potential data
points are very good; although this is generally true at all $\kappa$, in
this case we have picked a particularly good example 
(i.e., at $\kappa=0.1575$).     
 
Because of noisy data at large $R$ and also the exponential fall-off 
of the expectation value of the Wilson loop, the potential is poorly
determined at large distances. As already noted above,
the expression in Eq. (\ref{corr-pot}) ignores the running of the coupling 
$\alpha$ \cite{klassen}. Hence it is advisable to fit the potential in
as limited a range as possible \cite{edwards}.
On the other hand, determination of $\tilde{\sigma}$ which has the 
interpretation of the string tension at large $R$ becomes uncertain if $R$ is not taken large enough.
 
We fit the potential according to Eq. (\ref{corr-pot}) in the range 
$\{R_{\rm min},R_{\rm max}\}$ with
$R_{\rm min}\,=\, \sqrt{2},\, \sqrt{3},\, 
2 ,\, \sqrt{5},\, 2 \sqrt{2}$ and  
$R_{\rm max}\,=\, 4 \sqrt{2},\, 6,\, 3 \sqrt{5},\, 4 \sqrt{3},\,7$.

Fig. \ref{alpharmaxrmin} shows, at $\kappa\,=\,0.1575$ 
with APE smearing level 30,
the fit value of the parameter $\alpha$ for different values of
$R_{\rm min}$ and $R_{\rm max}$. The figure also shows 3 differnt data
sets corresponding to the 3 ranges 
$\left[T_{\rm min}, T_{\rm max}\right]\, =\, \left[3,\,4\right],\;
\left[3,\,5\right],\;\left[4,\,5\right] $ used for the evaluation of the 
potential by a single exponential fit. The figure clearly shows that
the errors for the fit with  
$\left[T_{\rm min}, T_{\rm max}\right]\, =\, \left[3,\,4\right]$ and
$R_{\rm min}\,=\,\sqrt{2}$ are the minimum. In addition, for this
choice, the data is also relatively independent of $R_{\rm max}$. For
these reasons we have chosen the fit range for the $V(R)$
evaluation as $\left[T_{\rm min}, T_{\rm max}\right]\, =\,
\left[3,\,4\right]$ and the fit range for fitting the potential to Eq. 
(\ref{corr-pot}) to be 
$\left[R_{\rm min}, R_{\rm max}\right]\, =\, \left[\sqrt{2},\,3\sqrt{5}\right]$.

\begin{figure}
\begin{center}
\includegraphics[width=.8\textwidth]
{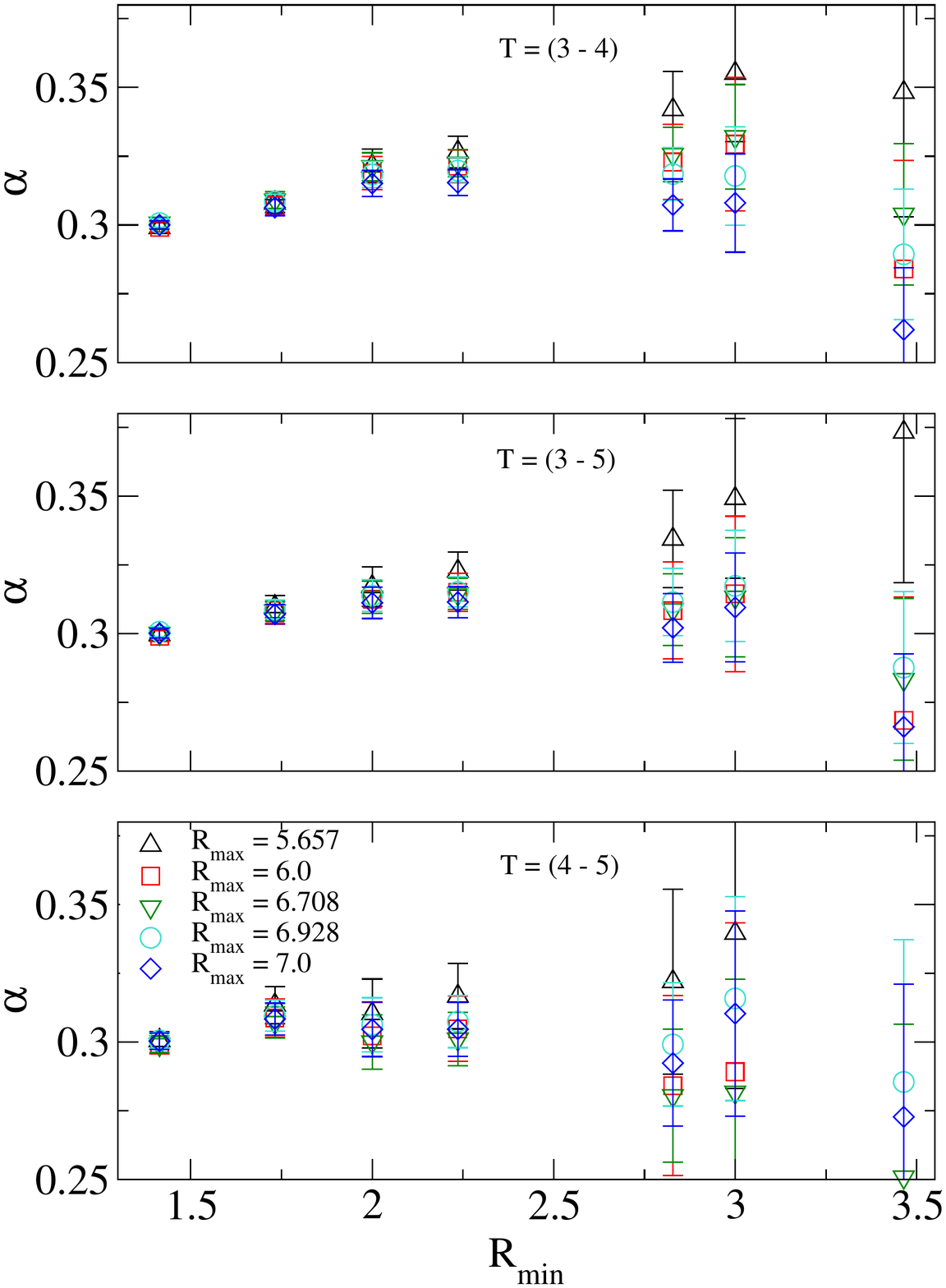}
\end{center}
\caption{The parameter $\alpha$ for $\kappa=$0.1575 as a function of
  $R_{\rm min}$ for three different $T$ ranges. For each $T$ range,
  the behavior for five choices of $R_{\rm max}$ are shown. }
\label{alphaRmin}
\end{figure}

The parameter $\alpha$ is always determined better at smaller $R_{\rm
  min}$. We also notice in Fig. \ref{alpharmaxrmin} that the value of
$\alpha$ grows larger as $R_{\rm min}$ increases, a fact consistent
with the expected running of $\alpha$. The results of the fit, i.e.,
the values of the parameters $\alpha$, $\tilde{\sigma}$ etc. are never 
fully invariant under the change of $R_{\rm min}$. As already pointed
above, $\alpha$ is sensitive to $R_{\rm min}$ and as a result the
other parameters of the correlated fit follow suit. The parameter $\alpha$ is only
approximately independent of  
$R_{\rm min}$ for $R_{\rm min}\geq 2$, but as far as any reliable determination
of $\alpha$ is concerned, the region $R_{\rm min}\geq 2$ is {\em not} 
reliable.  

Fig. \ref{alphaRmin} which plots $\alpha$ as a function of $R_{\rm min}$ for the three different fitting range $[T_{\rm min}, T_{\rm max}]$ clearly shows that $\alpha$ depends on $R_{\rm min}$ and the uncertainty in $\alpha$ increases with increasing $R_{\rm min}$. Based on this figure alone, perhaps the range  $[T_{\rm min}, T_{\rm max}]\,=\,[3,5]$ is equally good as the range $[3,4]$, but overall for all data the errors are smaller for  our chosen range $[3,4]$.  

At this point, let us also point out that $\alpha$ depends, albeit quite weakly, also on the
smearing level as more and more smearing progressively cuts out high
frequency modes. We have already explained, based on Fig. \ref{CR}, how
we have determined the level of smearing at each $\kappa$.

The results of the fit to Eq. (\ref{corr-pot}), as explained in
detail above, depend to some extent on various parameters related to
the fitting procedure 
($T_{\rm min},\, T_{\rm max}, \,R_{\rm min},\, R_{\rm max}$) and also
on the smearing level. {\em We have found that our final conclusions regarding the Sommer parameter, its dependence on the quark mass and the lattice scale are not very sensitive to change of these parameters of the data analysis and as a result our conclusions do not depend on the particular values used}.

The values of the fit parameters $aV_0$, $\alpha$, 
$\tilde{\sigma}\,=\,a^2\sigma$ and $\delta_{\rm ROT}$ are presented in
Table \ref{table1}. 

\begin{table}
\begin{tabular}{|l|l|l|l|l|l|l|l|}
\hline \hline
{$\kappa$} & {$aV_0$} & {$\alpha$} & {$a^2\sigma$}  & {$\delta_{ROT}$} &
      {$a/r_0$} & {$a/r_1$} & {$r_0/r_1$}\\
\hline
{0.156} & {0.6371(20)} & {0.2911(19)} & {5.748$\times 10^{-2}$(45)} & 
{0.3349(64)} & {0.2057(7)} & {0.2848(9)} & {1.3845(9)}\\
\hline 
{0.1565} & {0.6407(19)} & {0.2929(20)} & {5.363$\times 10^{-2}$(43)} & 
{0.3292(59)} & {0.1988(7)} & {0.2754(8)} & {1.3854(9)}\\
\hline
{0.15675} & {0.6457(17)} & {0.2958(16)} & {5.035$\times 10^{-2}$(34)} &
{0.3253(61)} & {0.1928(6)} & {0.2674(8)} & {1.3867(7)}\\
\hline
{0.157} & {0.6463(18)} & {0.2955(18)} & {4.892$\times 10^{-2}$(38)} & 
{0.3218(64)} & {0.1900(6)} & {0.2635(8)} & {1.3866(8)}\\
\hline
{0.15725} & {0.6507(16)} & {0.3010(17)} & {4.577$\times 10^{-2}$(34)} & 
{0.3291(58)} & {0.1842(6)} & {0.2559(8)} & {1.3892(8)}\\
\hline
{0.1575} & {0.6524(17)} & {0.3007(16)} & {4.416$\times 10^{-2}$(36)} & 
{0.3221(56)} & {0.1809(7)} & {0.2513(8)} & {1.3890(8)}\\
\hline
{0.15775} & {0.6559(16)} & {0.3022(16)} & {4.186$\times 10^{-2}$(37)} & 
{0.3184(55)} & {0.1762(7)} & {0.2449(9)} & {1.3898(8)}\\
\hline
{0.158} & {0.6560(15)} & {0.3013(16)} & {3.936$\times 10^{-2}$(34)} & 
{0.3173(53)} & {0.1708(7)} & {0.2374(9)} & {1.3894(8)}\\
\hline\hline
\end{tabular}
\caption{The four fit parameters of the static potential 
and the extracted values of  $a/r_0$, $a/r_1$ and $r_0/r_1$ for
  different $\kappa$.}
\label{table1}
\end{table}

\section{Sommer Scale}\label{sommer}


\begin{figure}
\begin{center}
\includegraphics[width=.8\textwidth]{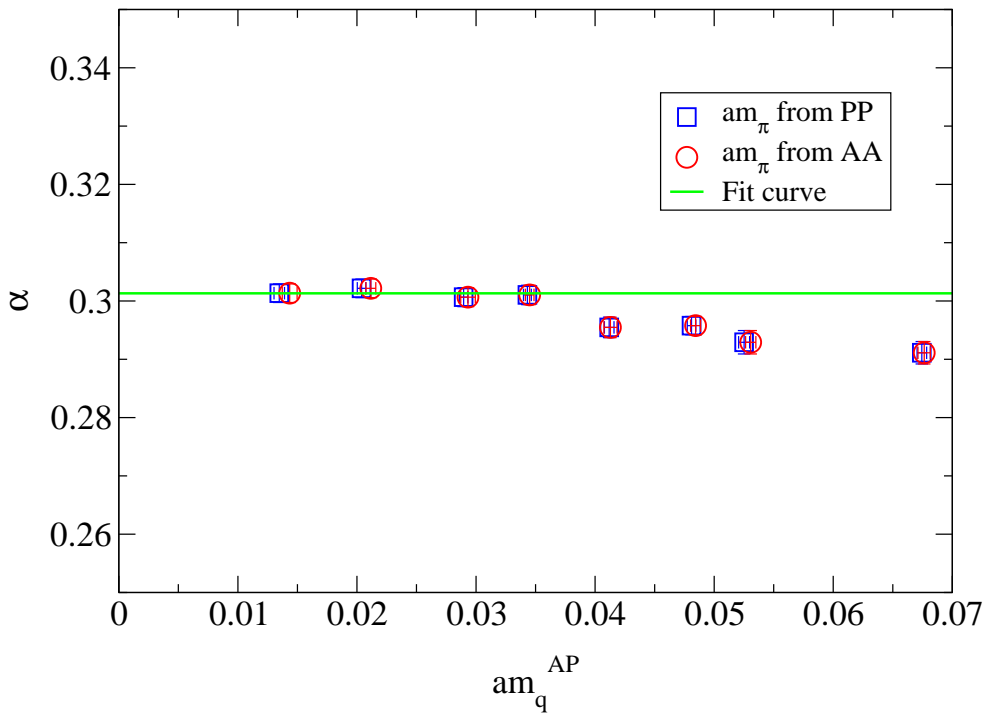}
\end{center}
\caption{The parameter $\alpha$ versus  two determinations of $a m_q^{\rm AP}$ (corresponding to whether the lattice pion mass used for the quark mass determination was taken from the PP or the AA correlator). The figure also shows
the straight line fit that incorporates the lowest four $am_q$
values. The notation $PP$, $AA$ and $AP$ is explained in Ref. \cite{paper0} and briefly in
Sec. \ref{simu}.}
\label{alpha}
\end{figure}
\begin{figure}
\begin{center}
\includegraphics[width=.8\textwidth]{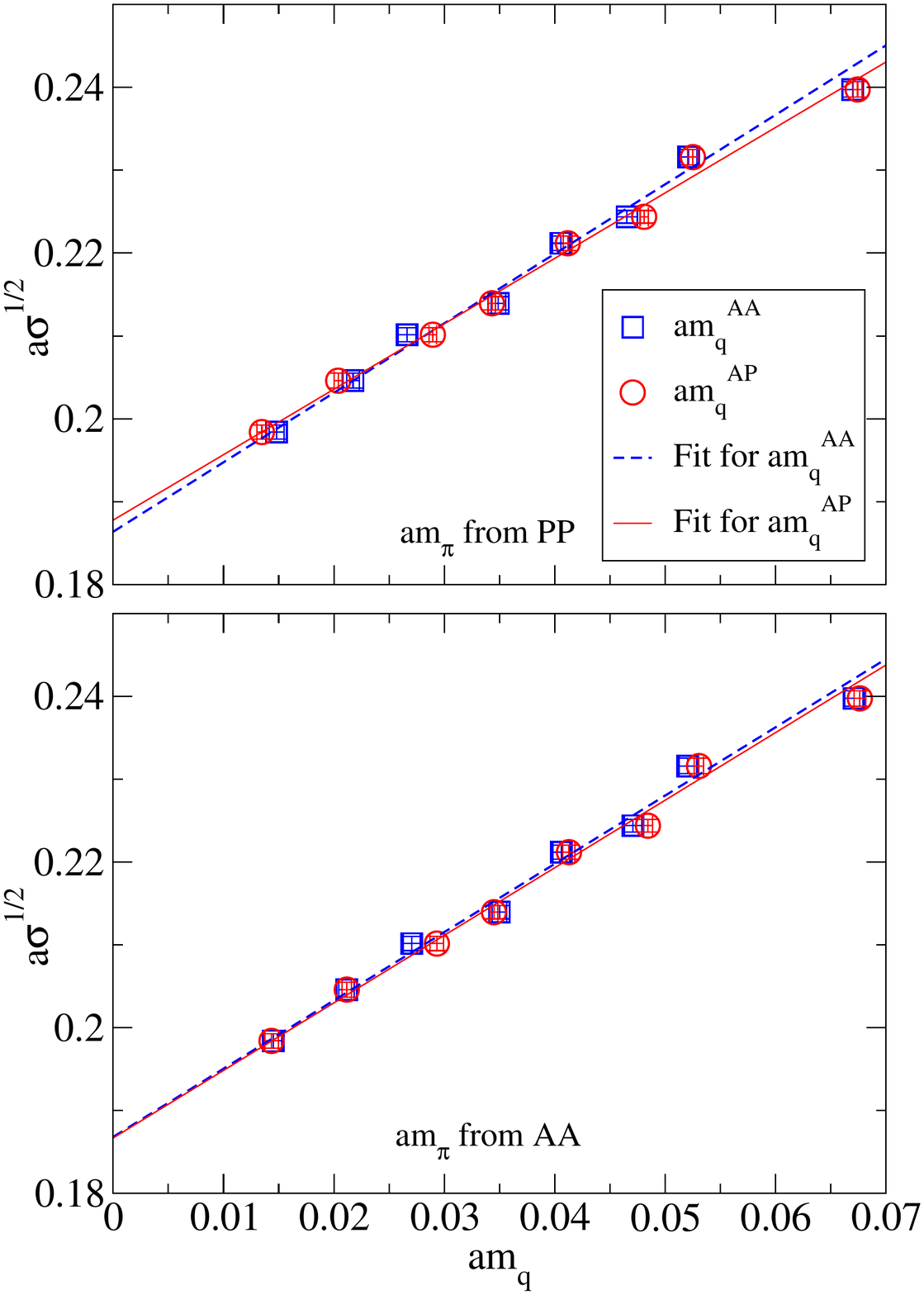}
\end{center}
\caption{The parameter $a {\sigma}^{1/2}$ vesus $a m_q$. The figure
  also shows the fits for the lowest five $am_q$ values as explained in the text.}
\label{sigma}
\end{figure}
\begin{figure}
\begin{center}
\includegraphics[width=.8\textwidth]{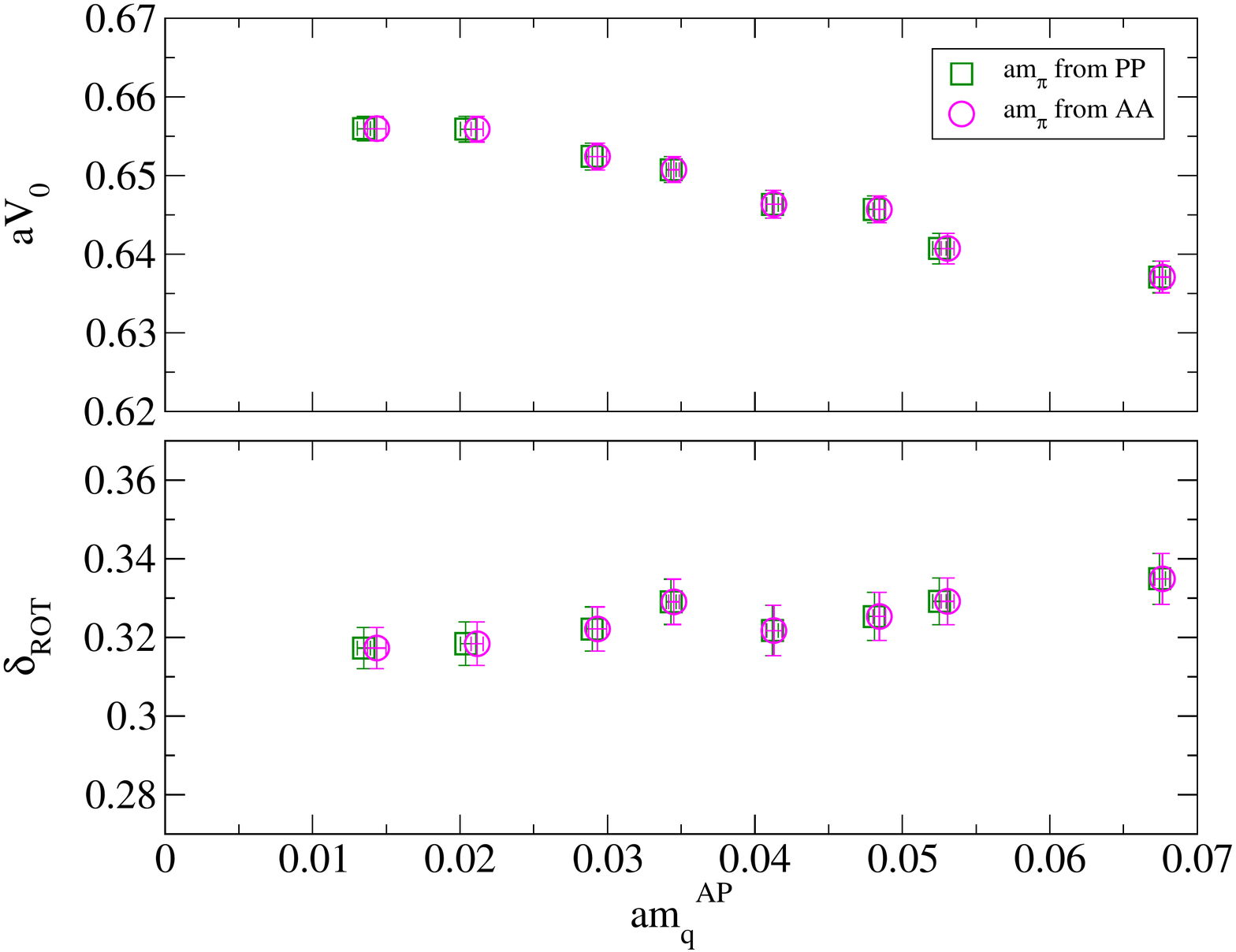}
\end{center}
\caption{The fit parameters $a V_0$ and $\delta_{\rm ROT}$ versus
  two different determinations of $am_q^{\rm AP}$ . }
\label{av0}
\end{figure}

After correcting for finite lattice effects, we compute the derivative
of the corrected (continuum) potential $aV_c$ given by the first three terms
of Eq. (\ref{corr-pot})
\be
R^2 \frac{d}{dR}aV_c = {\tilde \sigma}~ R^2 - \alpha~. \label{force}
\ee 

Phenomenological studies of the static  potentials yield \cite{sommer}
$ r^2 ~ \frac{d V}{d r}\mid_{r=r_{0}} ~ = ~1.65 $
at $r_0\simeq 0.49$ fm. The distance scale $r_0$ is known as the
Sommer scale. In general, there can be many such scales $r_c$ (in fm)
defined by 
\be
\left (r^2 ~ \frac{d V}{d r}\right )_{r=r_c} ~ = ~{\cal N}_c
\ee  
where ${\cal N}_c$ is a dimensionless number like 1.65 and $r_c$ is a
corresponding distance scale in fm obtained from the static
potential. In our case, using Eq. (\ref{force}), this means 
\be
\frac{a}{r_c} = \frac{1}{R_c} = 
\sqrt{ \frac{\tilde{\sigma}}{{\cal N}_c-\alpha}}
\label{abyrc}
\ee

Thus if we can determine the potential accurately in the intermediate 
range around $r_c$ fm, we can determine the lattice scale. 

Since we have used $R_{\rm min}=\sqrt{2}\sim1.4$ and $R_{\rm max}=3\sqrt{5}\sim6.7$ as the fit range for the static potential, the Sommer parameter $r_0\sim 0.49$ fm (corresponding to $R_0=r_0/a\sim6$) is barely within this range, we have also used another Sommer scale $r_1$ such that ${\cal N}_1\,=\,1.0$ \cite{milc1} (corresponding to $R_1=r_1/a\sim4.4$). However, we have not observed any noticeable improvement in the accuracy of the results obtained using the scale $r_1$,  because our fits describe the corrected potential data very accurately much beyond $R_{\rm max}$ as already observed in Sec. \ref{fit}. 

For any other potential-derived distance scales like $r_1$ (in fm) which are not
known phenomenologically as accurately as $r_0$, the strategy is to
compute the ratio $r_0/r_1$ accurately on the lattice and
determine $r_1$ from the ratio.    

\section{Dependence on $am_q$}\label{depamq}
Figs. \ref{alpha}, \ref{sigma} and 
\ref{av0} show respectively $\alpha$, 
$\tilde{\sigma}^{1/2}=a\sigma^{1/2}$, $aV_0$ and $\delta_{\rm ROT}$ versus
quark mass (derived from PCAC) in lattice units at all the eight
values of $\kappa$. These figures show that while $\alpha$ and $aV_0$
show only about $3\%$ change for the entire range of $am_q$ from 0.07
to below 0.015, $a\sigma^{1/2}$ go through a change of about $20\%$ in
the same range. Table \ref{table1} which has
$\tilde{\sigma}=a^2\sigma$ also shows about $30\%$ change in this
range. The values of the coefficient of the $1/R$-correction term, $\delta_{\rm
  ROT}$, shown in the lower part of
Fig. \ref{av0}, are quite close to the $\alpha$
values and are also similarly weakly-dependent on $am_q$.

\begin{figure}
\begin{center}
\includegraphics[width=.8\textwidth]{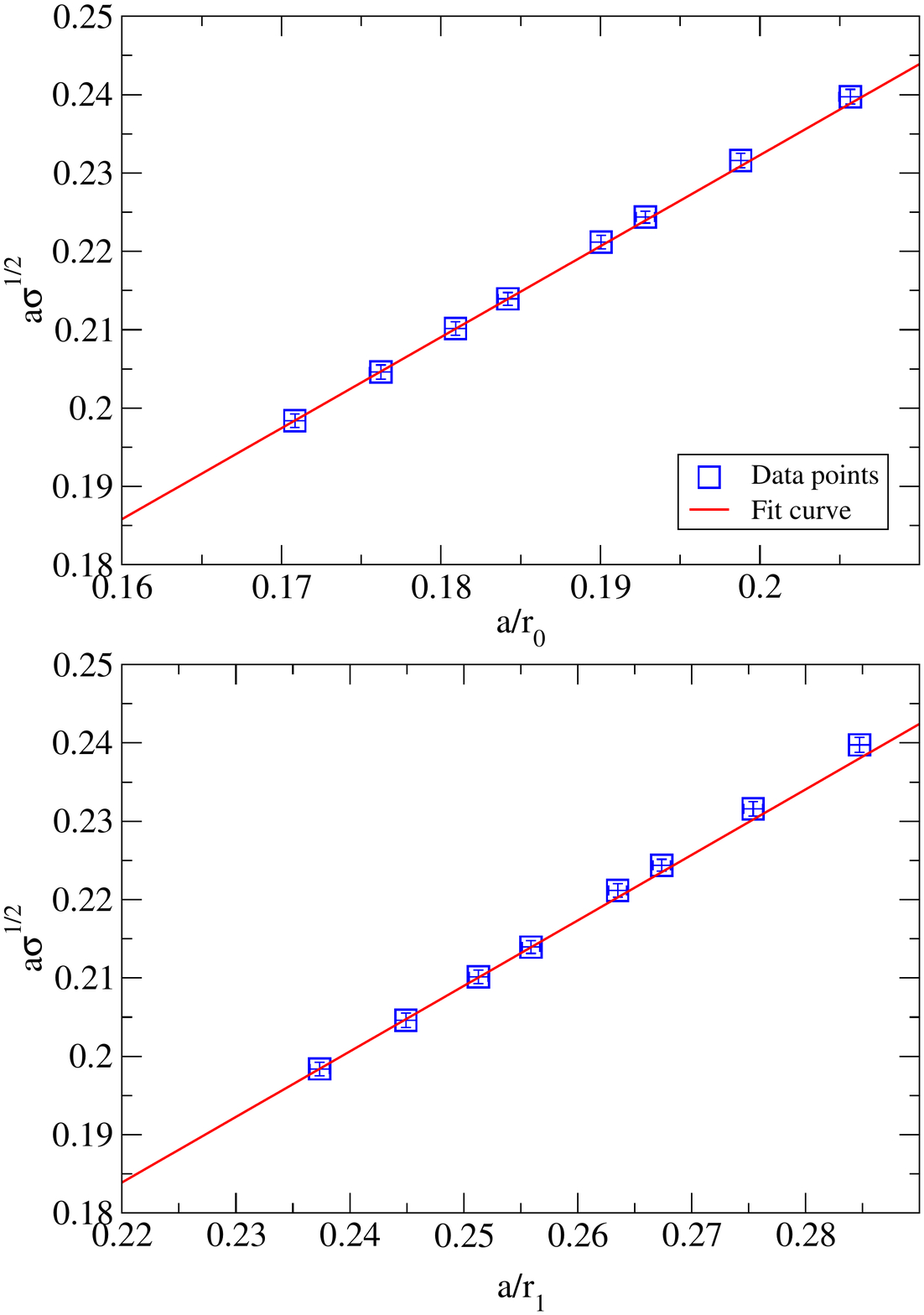}
\end{center}
\caption{The parameter $a\sigma^{1/2}$ versus $a/r_0$ and $a/r_1$. The
figures also show the fits with the lowest four data points as explained in the text. }
\label{K}
\end{figure}

Moreover, Fig. \ref{alpha} shows that, for small enough $am_q$ 
($am_q \lesssim 0.035$ with our data), $\alpha$ is independent of
$am_q$. That $\alpha$ is weakly dependent on $am_q$ and is roughly
constant within our accuracy is also seen in Fig. \ref{K} which shows that 
$a\sigma^{1/2}$ is proportional to $a/r_0$ and $a/r_1$ at least for
small enough $am_q$ and from Eq. (\ref{abyrc}) the proportionality constant 
is 
\be
K_c = \sqrt{{\cal N}_c-\alpha} 
\ee
with ${\cal N}_c\,=\,1.65$ and 1.0
respectively for $a/r_0$ and $a/r_1$. From Fig. \ref{K}, we compute 
$\alpha\,=\,0.3012(16)$ (from the $a/r_0$ dependence) and 
 $\alpha\,=\,0.3013(8)$ (from the $a/r_1$ dependence) for
$am_q \lesssim 0.035$. This coincides with the value
obtained from Fig. \ref{alpha}, viz.,
$\alpha\,=\,0.3013(8)$ again for $am_q \lesssim 0.035$.

Fig. \ref{sigma} shows $a\sigma^{1/2}$ to be linearly
dependent on $am_q$ with a positive intercept on the $a\sigma^{1/2}$
axis: 
\be
a\sigma^{1/2} = C_1+C_2 \;am_q \label{artsigma}
\ee
The fits use the lightest 5 quark masses corresponding to $am_q\lesssim 0.04$ and $C_1$ and $C_2$ are dimensionless constants. 

In the past, CP-PACS Collaboration tried non-linear chiral 
extrapolation for $a^2\sigma$ in Ref. \cite{aokichiralstring} and linear extrapolation of
$a\sigma^{1/2}$ later in Ref. \cite{alikhan}. 
We note that, UKQCD \cite{ukqcdalpha}, CP-PACS
\cite{aokistatic} 
and  SESAM and T$\chi$L  collaborations \cite{sesamtchil} 
have found $am_q$-{\em in}dependence of $\alpha$ in the small
$am_q$ region for Wilson or $O(a)$ improved Wilson fermions. 

\begin{figure}
\begin{center}
\includegraphics[width=.8\textwidth]{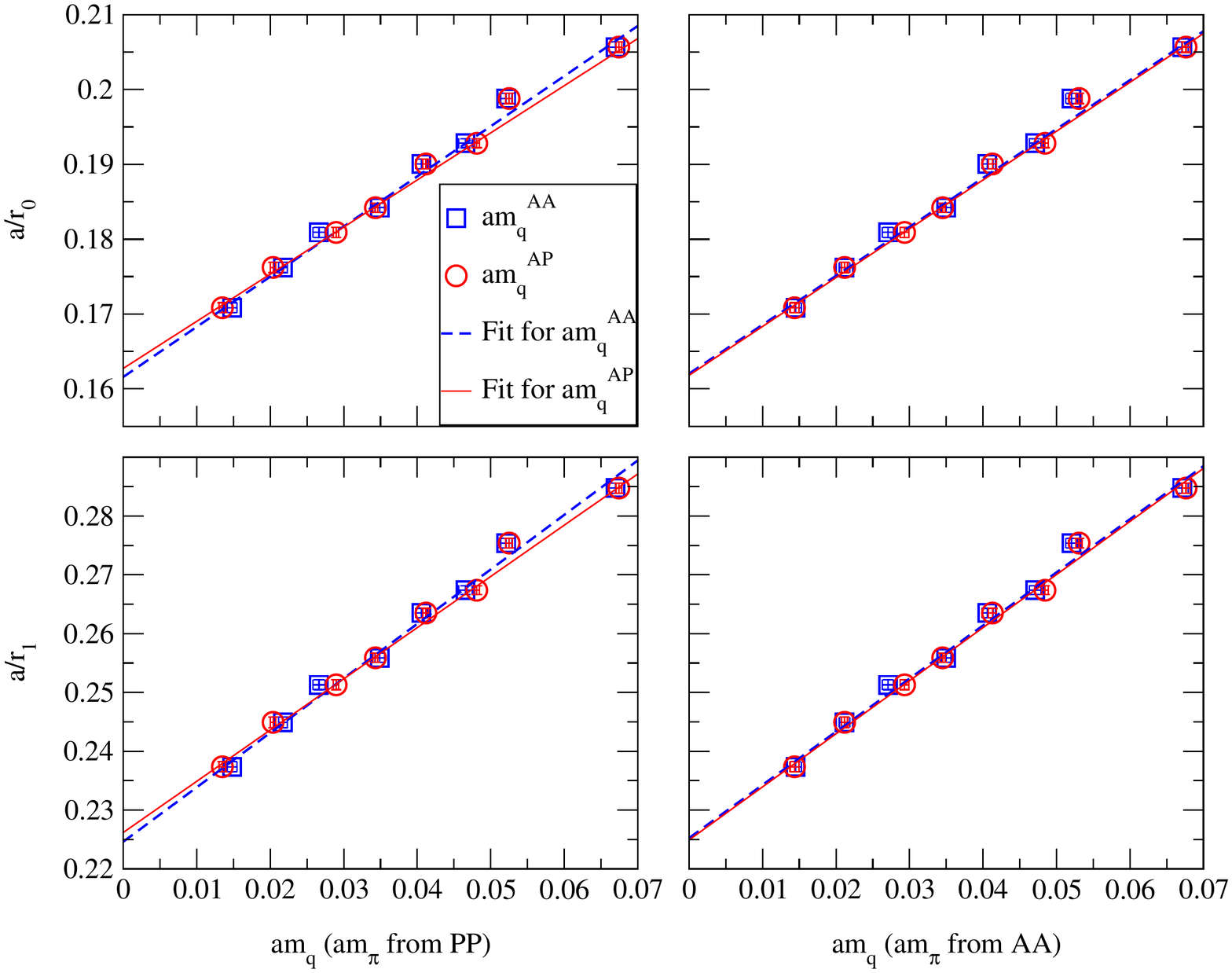}
\end{center}
\caption{Ratios $a/r_c$ versus  $a m_q$. The fits are done with the lowest five $am_q$ values.  }
\label{figabyrc}
\end{figure}
 
The dependence of $a/r_0$ and $a/r_1$ on $am_q$ is shown in Fig. \ref{figabyrc}. Consistent with the $am_q$-independence of $\alpha$ and the linear dependence of $a\sigma^{1/2}$, both $a/r_0$ and $ a/r_1$ depend linearly on $am_q$ at least for small enough $am_q$. The linear Fits
\be
\frac{a}{r_c} = A_c + B_c \; amq 
\label{rc_amq}
\ee 
are done for the lightest 4 quark masses (corresponding to $am_q\lesssim 0.035$). The dimensionless constants $A_c$ and $B_c$ are consistent with the $\alpha$ and the $a\sigma^{1/2}$ fits, i.e., $A_c\,=\, C_1/K_c$ and $B_c \,=\, C_2/K_c$ within our accuracy.

\section{Interpretation of the $am_q$ dependence of $a/r_c$}\label{interpret}

In the previous section, we have presented our numerical evidence of
$am_q$-dependence of $a/r_c$. For small $am_q\lesssim 0.035$, we have
shown that this dependence is linear and can be attributed to a
similar linear dependence of $a\sigma^{1/2}$ on $am_q$ while $\alpha$
does not appear to depend on $am_q$ within this range.

Given that our simulations are done with {\em unimproved} Wilson gauge and
fermion action, one may attribute {\em all} the observed $am_q$-dependence
to scaling violations \cite{sommercut}. However,  
{\em all} other serious investigations of lattice QCD with dynamical
quarks with a variety of {\em improved} gauge and fermion actions have
also observed a significant dependence of $a/r_c$ on $am_q$. For an
early summary of this dependence with improved and unimproved actions 
see Ref. \cite{aokistatus}. 
SESAM and T$\chi$L collaborations
\cite{sesamtchil} employed naive Wilson gauge and fermion actions and UKQCD
collaboration \cite{ukqcd} employed naive Wilson gauge and $O(a)$ improved
Wilson fermion actions whereas CP-PACS collaboration \cite{alikhan}
used both improved gauge and Wilson fermion actions. 
JLQCD Collaboration \cite{jlqcd} using Wilson
gauge action and $O(a)$ improved Wilson fermion action has also
observed this effect.  Same phenomenon was noticed in simulations with
standard and improved staggered fermions \cite{milc1,sg,milc2,milc3}. 
Recent simulations employing domain wall fermions \cite{domain},   
twisted mass fermions \cite{twisted} and overlap fermions
\cite{degrandtopo} 
have also observed the phenomenon. 

Obviously then, one cannot consider this purely as a cutoff effect. 
Some part of the quark mass
dependence of $a/r_c$, corresponding to relatively large $am_q$ is to
be considered scaling violations, but for small enough $am_q$ the
dependence should be accepted as a physical effect \cite{mcneile,milcstatus}.

Once one acknowledges that, for small enough $am_q$, the linear $am_q$ 
dependence of $a/r_c$ is a physical effect and {\em not} an artifact
of the cutoff, the natural choice consistent with a mass-independent
scheme is to assume that the scale $a$ is constant but $r_c$ changes with quark mass. To determine the scale $a$, one then needs to make a chiral extrapolation of $a/r_c$ to the physical point where estimates on $r_c$ are available from experimental data on the interquark potential in heavy-onium systems. In this scenario, the scale $a$ drops out of Eq. (\ref{rc_amq}) because it is the same for all values of the variables $1/r_c$ and $m_q$:
\be
\frac{1}{r_c} = {\cal A}_c + B_c \; m_q ~~~ {\rm with} ~~ A_c \, =\, a {\cal A}_c ~~
({\cal A}_c : ~{\rm a ~constant ~with ~dimensions ~of ~mass})
\label{rc_mq}
\ee 
showing $1/r_c$ to have a chiral behavior similar to $m_\rho$.

In this scenario, $\sigma^{1/2}$ also has a similar behavior (from Eq. (\ref{artsigma})):
\be
\sigma^{1/2} = {\cal C}_1 + C_2 \; m_q ~~~ {\rm with} ~~ C_1 \, =\, a {\cal C}_1 ~~
({\cal C}_1 : ~{\rm a ~constant ~with ~dimensions ~of ~mass})
\ee  

In our recent paper \cite{paper0} where we have investigated the pion and the rho masses and their decay constants as functions of the quark mass, we have assumed the above mass-independent scheme and the scale $a$ is independent of quark masses. 

In fact, application of chiral perturbation theory ($\chi PT$) is untenable if the scale $a$ is not taken independent of quark mass. The left hand side  of a $\chi PT$ equation always contains a physical quantity (like $m_\pi^2$) which is scale-independent. On the right hand side, there are low-energy constants and quark masses which are scale dependent. Obviously, a chiral extrapolation using such equations are {\em only} possible if all the quark masses are determined at the same scale $a$.  

We again stress that for the above interpretation of a mass-independent scale to work, $am_q$ has to be small enough. In our simulation with 2 degenerate flavors of sea quarks and with unimproved Wilson gauge and fermion actions we find that for $am_q\lesssim 0.035$, $\alpha$ is independent of $am_q$ and $a\sigma^{1/2}$ is linear in $am_q$, observations that lead to the linear $am_q$-dependence of $a/r_c$ (Eq. (\ref{rc_amq})). If, in addition, the scale $a$ is taken as independent of $m_q$, Eq. (\ref{rc_amq}) naturally evolves into Eq. (\ref{rc_mq}) which shows dependence of $1/r_c$ on quark mass $m_q$ as a physical effect.

One can consider the other extreme, i.e., $r_c$ independent of quark mass and the scale $a$ changing with quark mass. This is considered, for example,  in
Refs. \cite{ukqcd, ukqcd2002}, \cite{orthprd} and \cite{durr}. In this approach, the $am_q$ dependence shows up only as a scaling violation in the form \cite{schier1,schier2}
\be
\ln{\left(\frac{r_c}{a}\right)} = \sum a_k (\beta)^k + N \; am_q + ~~{\rm higher~ order~ scale~violations}, \label{logrcbya}
\ee
where $a_k$'s and $N$ are numbers. The $\beta$-dependence is a reflection of asymptotic scaling \cite{guagnelli}.
However, it needs to be pointed out that, although written above for $\ln{(r_c/a)}$, Eq. (\ref{logrcbya}) can be written down for any $\ln{(1/a\mu)}$ where $\mu$ is a hadronic scale like $m_\pi$ etc. Moreover, Eq. (\ref{logrcbya}) does {\em not} include any non-perturbative dependence of $\mu$ (e.g. quark mass dependence of hadronic masses). 

However, in this approach, it follows from Eq. (\ref{abyrc}) that for small enough $am_q$ the dimensionfull $\sigma^{1/2}$ is independent of the quark mass (because $\alpha$ is independent of $am_q$ for small $am_q$), something that apriori looks implausible because the string tension is nothing but the energy density of the field flux between the heavy quark-antiquark pair and is likely to depend on the sea quark masses. There are of course other problems associated with such a mass-dependent scheme, e.g., usage of $\chi pT$ and matching lattice scheme to mass-independent schemes like $\overline{MS}$.  

Our observation of $am_q$-independence of $\alpha$ for $am_q\lesssim 0.035$ indicates that scaling violations in our data are negligible for small $am_q$; however, Eq. (\ref{logrcbya}) suggests that scaling violations are always present even for small $am_q$. As a result we do not find support of the second scenario in our data and consequently pursue the first scenario where the scale $a$ is taken as a constant for a given $\beta$ and is determined from a chiral extrapolation of our $a/r_c$ data to the physical point. 

\section{Chiral extrapolation}\label{chiral} 
\begin{figure}
\begin{center}
\includegraphics[width=.8\textwidth]{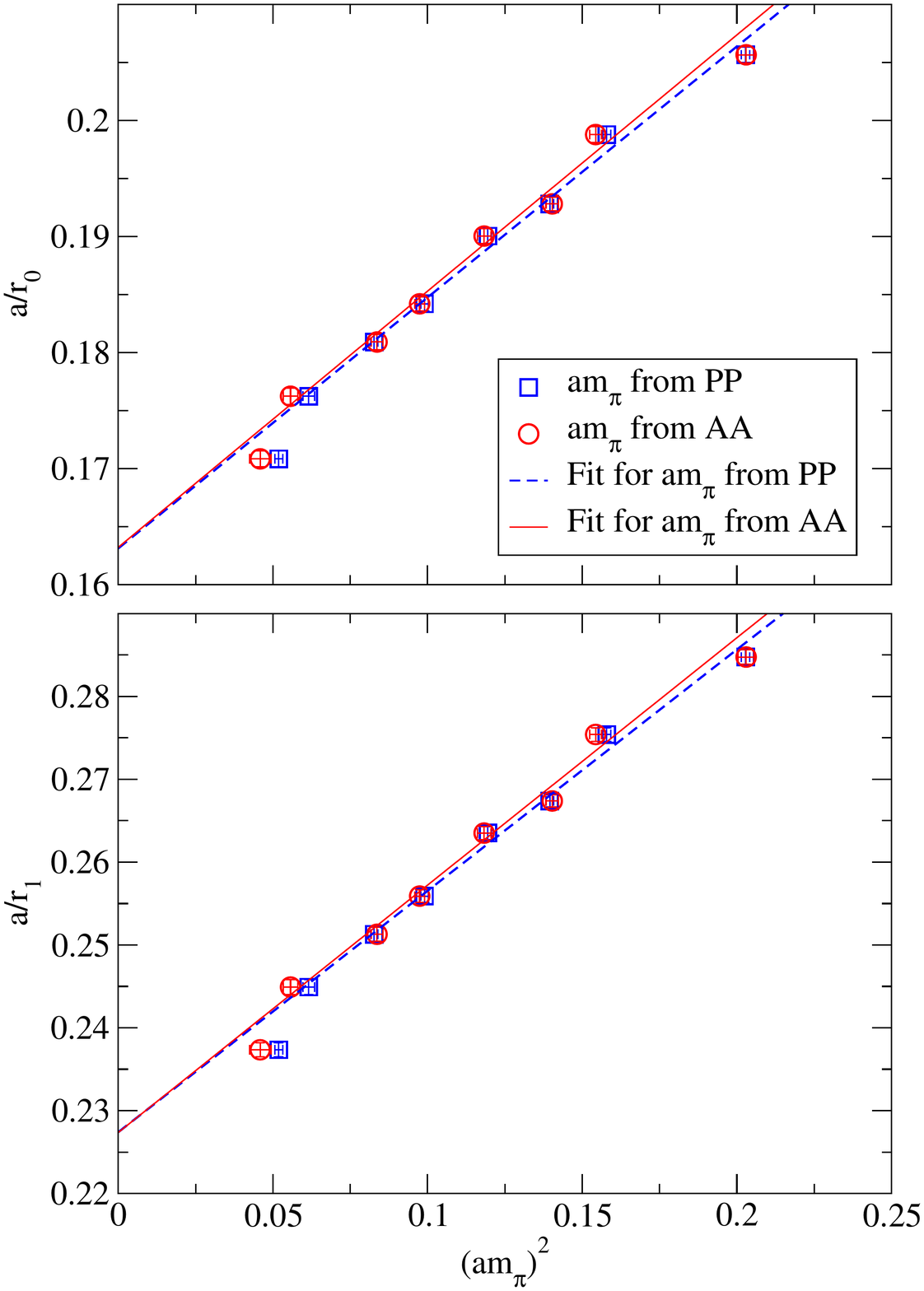}
\end{center}
\caption{Ratios $a/r_c$ versus  ${(a m_\pi)}^2$. The fits are done with four lowest
pion masses free from finite size effect.}
\label{abyrc_ampisq}
\end{figure}

\begin{figure}
\begin{center}
\includegraphics[width=.8\textwidth]{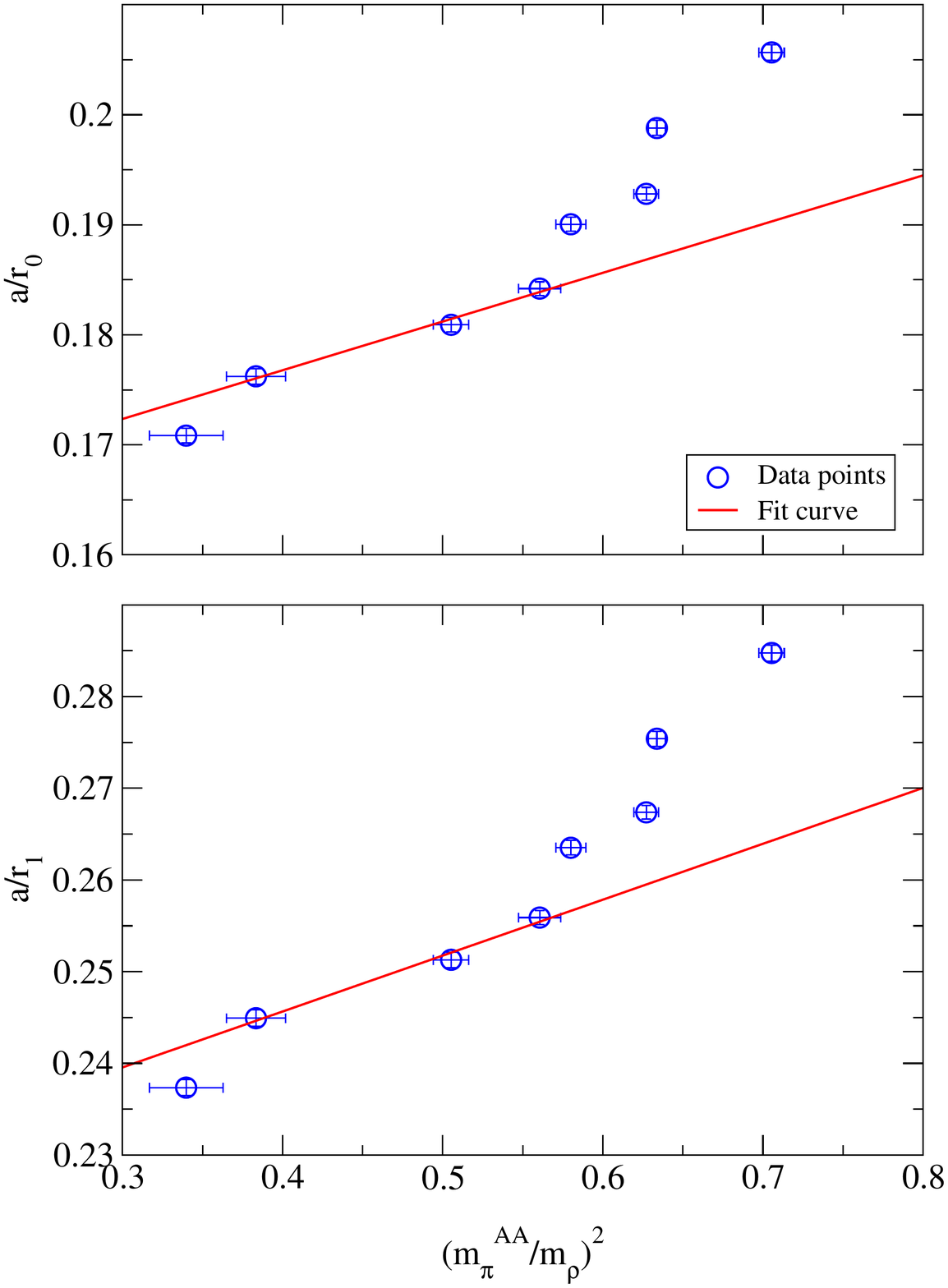}
\end{center}
\caption{Ratios $a/r_c$ versus
  $(m_\pi/m_\rho)^2$. The fits are done with the three lowest $(m_\pi/m_\rho)^2$ free of finite size effect.}
\label{abrcmpmrsq}
\end{figure}

\begin{figure}
\begin{center}
\includegraphics[width=.8\textwidth]{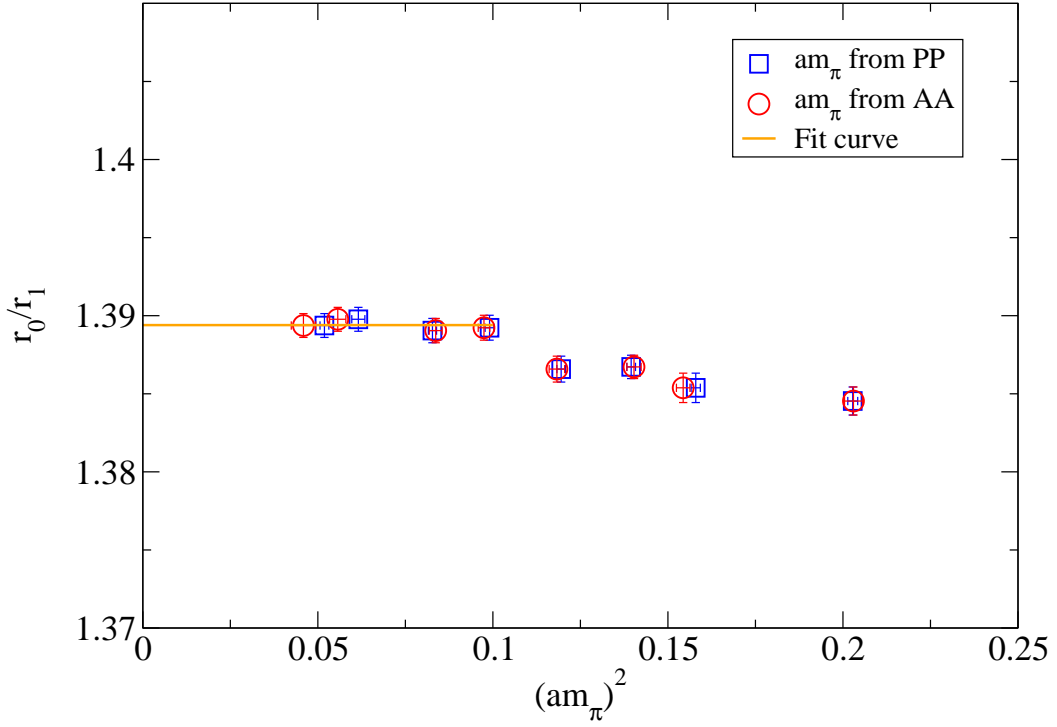}
\end{center}
\caption{The ratio $r_0/r_1$ versus  $(a m_\pi)^2$. The
  fit is done with the four lowest $(a m_\pi)^2$ values.}
\label{r0byr1}
\end{figure}

\begin{table}
\begin{center}
\begin{tabular}{|c|c|c|c|c|c|c|c|c|}
\hline \hline
&\multicolumn{4}{c|}{Chiral limit of $a/r_0$}&\multicolumn{4}{c|}{Chiral limit of $a/r_1$}\\
\cline{2-9}
{Extrapolation}& \multicolumn{2}{c|}{$am_\pi$ from PP} & \multicolumn{2}{c|}{$am_\pi$ from AA}
& \multicolumn{2}{c|}{$am_\pi$ from PP} & \multicolumn{2}{c|}{$am_\pi$ from AA}\\
\cline{2-9}
& {$am_q^{AA} $ } & {$am_q^{AP} $} &{$a m_q^{AA}$ } &{$am_q^{AP} $ } 
&{$ am_q^{AA}$ } &{$ am_q^{AP}$ } &{$ am_q^{AA}$ } & {$am_q^{AP} $ } \\
\hline
$a/r_{c}=A_c + B_c a m_q$ & {\bf 0.1616(13)} & {\bf 0.1627(10)} &
{\bf  0.1620(13)} & {\bf 0.1618(12)} & {\bf 0.2246(18)} & {\bf 0.2262(13)} 
 & {\bf 0.2253(17)} & {\bf 0.2249(15)}\\
\hline
$K_c=\sqrt{{\cal N}_c - \alpha}$ & & & & & & & & \\
$a \sigma^{1/2}= C_1+C_2 am_q $& 0.1605(12)& 0.1617(10)
& 0.1609(12)& 0.1607(11)& 0.2230(17)& 0.2246(13)
& 0.2235(17)& 0.2233(14)\\
$a/r_{c}=\frac{ C_1}{K_c} + \frac{ C_2}{K_c} a m_q$ 
& {} &{} &{} &{} & {}  & {} & {}& {}\\
\hline
{$a/r_{c}=P_c+ Q_c (am_\pi)^2  $} & \multicolumn{2}{c|}{0.1631(16)} 
& \multicolumn{2}{c|}{0.1632(16)}
& \multicolumn{2}{c|}{0.2274(20)} 
& \multicolumn{2}{c|}{0.2274(20)}\\
\hline
{$a/r_{c}=D_c+ E_c (m_\pi/m_\rho)^2  $} & \multicolumn{2}{c|}{-} 
& \multicolumn{2}{c|}{0.1591(37)}
& \multicolumn{2}{c|}{-} 
& \multicolumn{2}{c|}{0.2213(49)}\\
\hline
\hline
\end{tabular}
\caption{The values of $a/r_c$ extracted in the chiral limit
  using four different extrapolations. No entries for the extrapolation with $(m_\pi/m_\rho)^2$ when $m_\pi$ is determined from the PP correlator (because of finite size effects on the PP correlator at the largest two $\kappa$ values, as expalined in text).}   
\label{tababyrc}
\end{center}
\end{table}

On the lattice $am_q$ is the best regulator of chiral symmetry breaking and as such it is the best parameter to use for chiral extrapolation of $a/r_c$. However, it is not a good parameter for an approach to the physical point because apriori one does not know the value of $am_q$ at the physical point although it is very close to the chiral limit.

For the chiral extrapolation of $a/r_c$ to the physical point, we have used $(am_\pi)^2$. Fig. \ref{abyrc_ampisq} plots both $a/r_0$ and $ a/r_1$ versus $(am_\pi)^2$. Firstly, we emphasize that we prefer $(am_\pi)^2$ rather than $(r_cm_\pi)^2$, because $r_c$ has its own quark mass dependence. Secondly, for the fit we stick to the linear part of the dependence corresponding to small $(am_\pi)^2$. No higher powers of $(am_\pi)^2$ is entertained to fit {\em all} data because as much as possible we want to stay away from data points which may have some scaling violations. Unfortunately, for a quantity like $(am_\pi)^2$ there is significant finite size effect at the smallest values with our lattice volumes. According to the findings of Ref. \cite{paper0} we dropped the lowest two values when fitting with $(am_\pi)^2$ determined from PP correlators, and dropped only the lowest point when fitting with $(am_\pi)^2$ determined from AA correlators.

We obtain the scale $a$ by solving the quadratic equation in $a$:
\be
\frac{a}{r_c^{\rm ph}} = P_c + Q_c\left (am_\pi^{\rm ph}\right )^2  
\label{quadeq}
\ee
where $P_c$ and $Q_c$ are constants, and $r_c^{\rm ph}$ and $m_\pi^{\rm ph}$ the values at the physical point.

We acknowledge that there is some unavoidable mistake made by using $(am_\pi)^2$ (as opposed to using $am_q$) for the chiral extrapolation because in the chiral region the relation between $(am_\pi)^2$ and $am_q$ is linear only in the so-called leading order. We feel that using $(am_\pi)^2$ is still better than using $(r_cm_\pi)^2$ for reasons stated above and definitely better than using $(m_\pi/m_\rho)^2$ which we also use for a rough estimate of the chiral extrapolation in Fig. \ref{abrcmpmrsq}.   The squared ratio $(m_\pi/m_\rho)^2$ is generally taken as an estimate of the quark mass, but it is not linear in quark mass for any appreciable range of quark mass. For small enough quark masses, Fig. \ref{abrcmpmrsq} shows approximate linear behavior for the smallest masses. We have done the scale determination using chiral extrapolation of $a/r_c$ with respect to $(m_\pi/m_\rho)^2$ with pion masses determined only from the AA correlator because only in that case we have three data points (excluding the lightest masses at $\kappa=0.158$ but including the data at $\kappa=0.15775$) for a straight line fit. As pointed out by our earlier work \cite{paper0}, there are significantly more finite size effects on the pion mass determined from the PP correlator and we have to drop the lightest two masses (corresponding to $\kappa=0.158$ and $0.15775$) and as a consequence would be left with only two points for a linear chiral extrapolation in this case. Hence we do the chiral extrapolation of $a/r_c$ in dependence of $(m_\pi/m_\rho)^2$ when the pion mass is determined only from the AA correlator.  

In Table \ref{tababyrc} we show the $a/r_c$ values at the {\em chiral limit} obtained with extrapolations done using $am_q$ (first data row), $(am_\pi)^2$ (third data row) and $(m_\pi/m_\rho)^2$ (fourth data row). The second data row contains the chiral limit values of $a/r_c$ obtained from the individual limits of $\alpha$ and $a\sigma^{1/2}$. The first and the second row values are consistent with each other showing that our inference of $\alpha$ being almost independent of $am_q$ and $a\sigma^{1/2}$ linear in $am_q$ is correct. Comparison of the values in the first and the third data rows shows that the central values are about $1\%$  off and they are consistent with each other within statistical errors. These consistency checks give credibility to the chiral extrapolation of $a/r_c$ with respect to $(am_\pi)^2$. Only the fourth data row containing extrapolated values using $(m_\pi/m_\rho)^2$ shows a deviation of about $3\%$ from the values in the first data row and also exhibit significantly larger statistical errors. The first data row is in bold font to emphasize that the data entries in this row have the most reliable chiral limits.  

\begin{table}
\begin{center}
\begin{tabular}{|c|c|c|c|c|c|c|}
\hline \hline
& \multicolumn{4}{c|}{$ (a/r_{1})/(a/r_{0})$} & {$r_0/r_1$}&\\
\cline{2-5}
& \multicolumn{2}{c|}{$am_\pi$ from PP} & \multicolumn{2}{c|}{$am_\pi$ from AA}
& versus&{$r_0/r_1=K_0/K_1$} \\
\cline{2-5}
& {$a m_q^{AA}$} & {$a m_q^{AP}$} &{ $a m_q^{AA}$ } &{$am_q^{AP} $ }   &
      {$(a m_\pi)^2$} &\\
\hline
$r_0/r_1$ & {1.3900(16)} &{1.3899(14)} &{1.3901(15)} 
&{1.3899(14)} & {\bf 1.3894(7)}  & {1.3894(4)} \\
\hline
\hline
\end{tabular}
\caption{The  ratio $r_0/r_1$ in the chiral limit extracted
  using different methods.}
\label{tabr0byr1}
\end{center}
\end{table}

\begin{table}
\begin{center}
\begin{tabular}{|c|c|c|c|c|c|c|c|c|}
\hline \hline
&\multicolumn{4}{c|}{$a/r_0$ fit}&\multicolumn{4}{c|}{$a/r_1$ fit}\\
\cline{2-9}
{Extrapolation}& \multicolumn{2}{c|}{$am_\pi$ from PP} & \multicolumn{2}{c|}{$am_\pi$ from AA}
& \multicolumn{2}{c|}{$am_\pi$ from PP} & \multicolumn{2}{c|}{$am_\pi$ from AA}\\
\cline{2-9}
{to the physical point}
& {$a $ (fm) } & {$a^{-1} $ (GeV)} &{$a$ (fm)} &{$ a^{-1} $ (GeV) } 
&{$ a$ (fm)} &{$ a^{-1}$ (GeV)} &{$a $ (fm)} & { $ a^{-1} $ (GeV)} \\
\hline
$a/r_c=P_c + Q_c (a m_\pi)^2$ & {\bf 0.08027(77)} &{\bf 2.458(23)} 
&{\bf 0.08032(76)} &{\bf 2.457(23)} 
& {\bf 0.08053(70)}  & {\bf 2.450(21)} & {\bf 0.08053(71)}& {\bf 2.450(22)}\\
\hline\
$a/r_{c}=D_c + E_c  (m_\pi/m_\rho)^2$ 
& {--} &{--} &{0.07865(170)} &{2.509(54)} 
& {--}  & {--} & {0.07873(164)}& {2.506(52)}\\
\hline
\hline
\end{tabular}
\caption{The lattice scale $a$ (fm) and $a^{-1}$ (GeV) obtained with two different extrapolations of $a/r_c$ to the physical point. }
\label{afit}
\end{center}
\end{table}

\begin{table}
\begin{center}
\begin{tabular}{|c|c|c|c|c|}
\hline \hline
Extrapolation to & \multicolumn{2}{c|}{$am_\pi$ from PP} & \multicolumn{2}{c|}{$am_\pi$ from AA}   \\
\cline{2-5}
{the physical point}
& {$ a$ (fm)} & {$ a^{-1}$ (GeV)} &{ $ a$ (fm) } &{$a^{-1} $ (GeV) }     \\
\hline
$am_\rho=F_1+F_2(am_\pi)^2$ & {0.07932(135)} &{2.488(41)} &{0.07995(195)} &{2.468(60)}  \\
\hline
\hline
\end{tabular}
\caption{The lattice scale $a$ (fm) and $a^{-1}$ (GeV) obtained with extrapolation
 of $am_\rho$ to the physical point.}\label{afit_mrho}
\end{center}
\end{table}

Fig. \ref{r0byr1} plotted against $(am_\pi)^2$ shows that for $(am_\pi)^2\lesssim 0.1$ or $am_q\lesssim 0.035$ the ratio $r_0/r_1$ is independent of these quantities. The fitted constant value in the figure is 1.3894(7) (shown in bold in Table \ref{tabr0byr1} to indicate that this is the value actually used to determine $r_1$). This value is absolutely consistent with the ratio of the chiral limits $a/r_1$ to $a/r_0$ (extrapolated with different quark mass evaluations) and also with the ratio $K_0/K_1\,=\,((1.65-\alpha)/(1.00-\alpha))^{1/2}$ (with the value of the fitted $\alpha$ put in from Fig. \ref{alpha}), as shown in Table \ref{tabr0byr1}.

We have used $r_0\,=\,0.49$ fm in our analysis. We are aware that a few lattice groups have calculated $r_0$ from the low level splittings of the bottomonium system and those values are a few percent lower than the standard value used in this paper. In any case we have done our simulation with 2 degenerate sea quarks only and since we have taken the view in this paper that $r_c$ changes with quark mass, it is conceivable that $r_c$ may change somewhat when the number of flavor is changed. We take the viewpoint that given that
$r_0$ may have some uncertainties, we try to minimize all other uncertainties regarding the scale determination.   

Given the $r_0/r_1$ value at the physical point, with $r_0\,=\,0.49$ fm, we get $r_1\,=\, 0.3527(2) $ fm. We can now plug in the values of $r_0$ and $r_1$ respectively in the values of $a/r_0$ and $a/r_1$ at the physical point obtained from the various extrapolations of $a/r_c$ and get the scale $a$. Table \ref{afit} lists the values of $a$ in fm and $a^{-1}$ in GeV obtained with the two methods employed. We notice that the scales obtained from $(am_\pi)^2$-extrapolation has very accurate values with less than $1\%$ errors (emphasized by bold fonts in the table) while those obtained from $(m_\pi/m_\rho)^2$ has about $2\%$ errors. Within error bars the values are consistent. 

\begin{figure}
\begin{center}
\includegraphics[width=.8\textwidth]{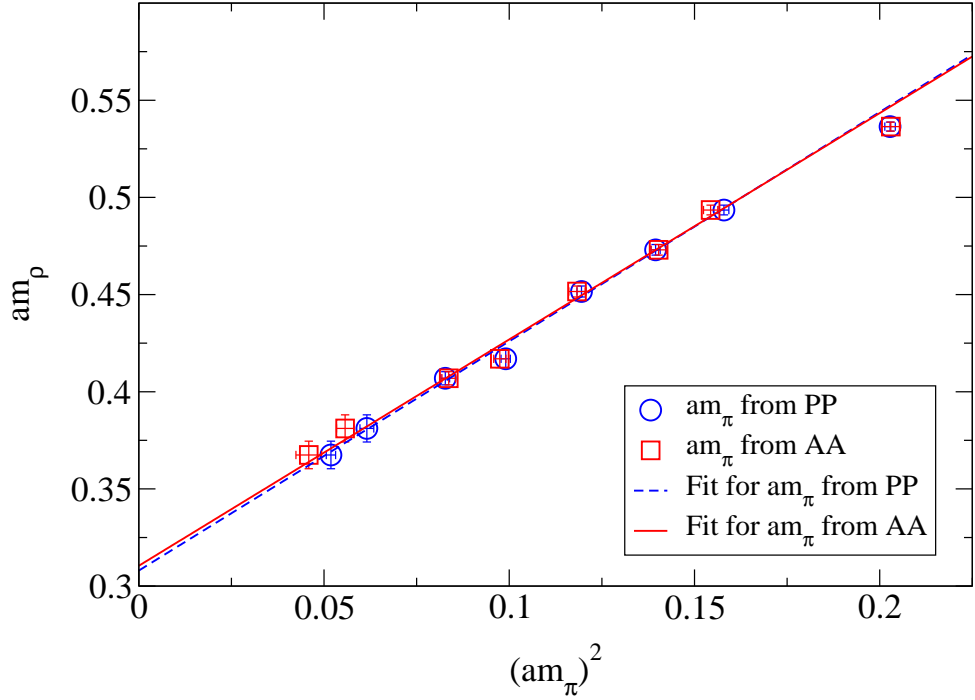}
\end{center}
\caption{$a m_\rho$ vs $(a m_\pi)^2$. The fits are done with the lowest five pion masses free of finite size effects}
\label{rhovspisq}
\end{figure}

In our earlier paper \cite{paper0} we computed the rho meson mass in lattice units and it was shown to have a linear behavior in $am_q$ with a positive intercept at $am_q=0$. In Fig. \ref{rhovspisq} we plot $am_\rho$ as a function of $(am_\pi)^2$. The data is well fit by a linear ansatz:
\be
am_\rho = F_1 + F_2 (am_\pi)^2.
\ee 
The smallest two and the largest pion masses are excluded from the fit for the case of the pion mass determined from the PP correlator while for the fit with pion mass determined from the AA correlator we have excluded the largest two and the lightest pion mass. These exclusions are due to possible finite size effects on the smallest pion masses and possibility of being outside the chiral regime for the larger pion masses. The value of the intercept at $(am_\pi)^2=0$ is consistent with the intercept at $am_q=0$ in Ref. \cite{paper0}. Again, the fit can be looked upon as a quadratic equation in the scale $a$ (in fm) at the physical point while the pion mass $m_\pi^{\rm ph}$ and the rho mass $m_\rho^{\rm ph}$ are entered in fm$^{-1}$. The scales obtained by solving the equation are independent of the static potential and scales determined therefrom and are listed in Table \ref{afit_mrho}. Although with relatively larger errors ($\sim 2-2.5\%$), these scales are very close to our very accurate evaluations using the $a/r_c$ extrapolations given in Table \ref{afit}.

\section{Estimate of $\sigma^{1/2}$}\label{sigmad}
\begin{figure}
\begin{center}
\includegraphics[width=.8\textwidth]{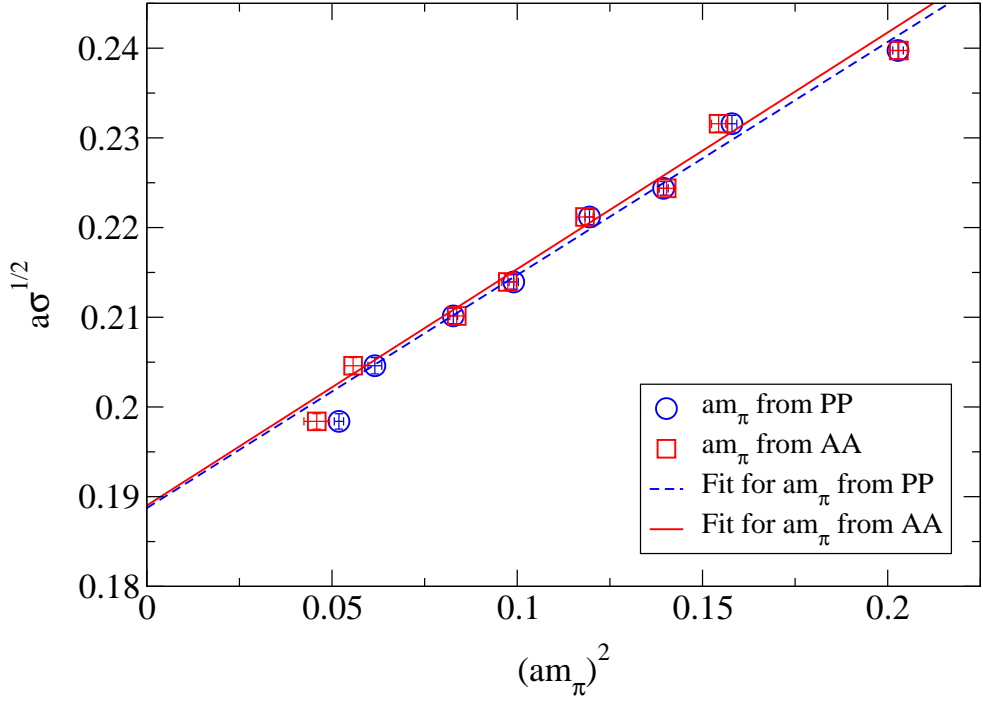}
\end{center}
\caption{$a\sqrt{\sigma}$ vs $(a m_\pi)^2$. The fits are done with the four lowest pion masses free of finite size effects.}.
\label{sqrtsigmavspisq}
\end{figure}

We can now make estimates of the parameter $\sigma^{1/2}$ in physical dimensions by chirally extrapolating $a\sigma^{1/2}$ with 
$(am_\pi)^2$ using the scale determined by our accurate determinations from $a/r_c$ extrapolations (first data row of Table \ref{afit}) and the pion mass $m_\pi^{\rm ph}$ at the physical point:
\be
a(\sigma^{\rm ph})^{1/2} = G_1 + G_2 \left (a m_\pi^{\rm ph}\right )^2.
 \label{sigph}
 \ee
The data along with the fits to determine the constants $G_1$ and 
$G_2$ are shown in Fig. \ref{sqrtsigmavspisq}. The points included for the fits are similar to the $am_\rho$ - $(am_\pi)^2$ fits discussed before.  From Eq. (\ref{sigph}) we obtain four values of  $\sigma^{1/2}$ (dropping the superscript indicating the value at the physical point) corresponding to two evaluations of pion masses from the PP and the AA correlator and two values of the scale  $a$ from $a/r_0$ and $a/r_1$ extrapolations. All these four values are extremely close to each other. We present the average and quote the largest error of the four:   
\be
\sigma^{1/2} = 465.5 \pm 1.4 ~{\rm MeV}.
\label{sigval1}
\ee  

\begin{figure}
\begin{center}
\includegraphics[width=.8\textwidth]{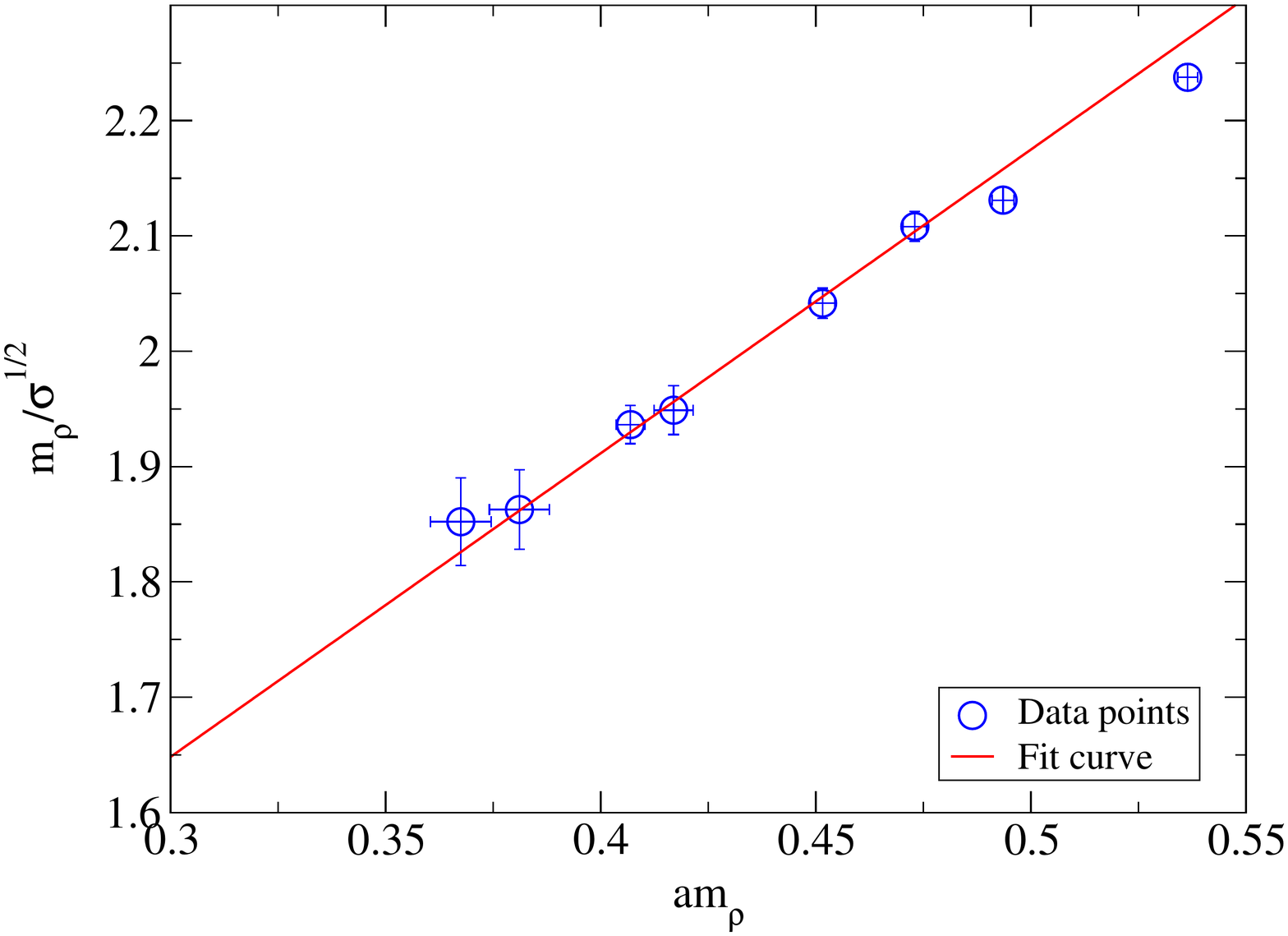}
\end{center}
\caption{The ratio $m_\rho/\sqrt{\sigma}$ versus $a m_\rho$. The fit is done with the lowest five data points excluding the lowest point.}
\label{rhosqrtsigratio}
\end{figure}

Similarly, using the scale determined independent of the static potential (from $am_\rho$ - $(am_\pi)^2$ fits), $\sigma^{1/2}$ can be independently determined from the linear behavior of the ratio 
$m_\rho/\sigma^{1/2}$ with $am_\rho$ (shown in Fig. \ref{rhosqrtsigratio}). These determinations of $\sigma^{1/2}$ have somewhat larger statistical errors than above. We present the average with the largest error:
\be
\sigma^{1/2} = 460.9\pm 9.3 ~{\rm MeV}.
\label{sigval2}
\ee

Although the statistical error in Eq. (\ref{sigval1}) is surprisingly very small, obviously it does not represent all the errors associated with the evaluation of $\sigma^{1/2}$ as the second evaluation given in Eq. (\ref{sigval2}) shows a significant systematic deviation of the central values.

\section{Implications from weak coupling perturbation theory}\label{impli}

Once we have determined the parameters of the static potential, it is
interesting to compare the extracted parameters with those expected
from weak coupling perturbation theory, wherever appropriate. For
example, one would like to see how  the extracted self energy
quantitatively differs from that calculated in one loop (tadpole
improved \cite{lm}) lattice perturbation theory. Moreover, from the
non-perturbatively extracted average value of the plaquette, one can
extract the strong coupling constant $\alpha_{\rm v}$ at the scale $3.41/a$
according to the Lepage-Mackenzie scheme \cite{lm}.

In weak coupling perturbation theory, the expectation value of the
Wilson loop is given by
\be
{\rm Limit_{T \rightarrow \infty}} (-) ~ \frac{1}{T} ~{\rm ln} \langle W(R,T)
  \rangle ~=~ V(R)
\ee
where
\be
V(R) = V_{\rm coul}~+~ V_{\rm self}~. 
\ee
The static Coulomb potential 
\be
V_{\rm coul}~=~ - C_F~\frac{\alpha_s}{R}
\ee
and $V_{\rm self}$ is the static source self energy. 
\subsection{Static source self energy}
In the continuum, to lowest order, 
\be
V_{\rm self}~=~ C_F~4 \pi \alpha_s ~\int \frac{d^3q}{(2 \pi)^3}~
\frac{1}{{\bf q}^2} 
\ee
On the lattice
\be
a V_{\rm self} = C_F~4 \pi \alpha_s ~ \frac{1}{L^3}~ \sum_{q_i \neq
  0}~ \frac{1}{\sum_i  \sin^2 a q_i/2}~.
\ee
Using $\beta=5.6$ and $\frac{4 \pi}{L^3}~ \sum\limits_{q_i \neq
  0}~ \frac{1}{\sum_i { \sin^2} a q_i/2}$ = 2.9987 for a
$16^3$ lattice, we get $a V_{\rm self}$=0.3409. Incorporating tadpole 
improvement utilising the average value of the plaqutte ($\Box_{\rm
  av}$), 
$ g^2 \rightarrow {\tilde g}^2 = \frac{g^2}{\langle {\Box} \rangle}= 
g^2/0.5744$, we get
 $a V_{\rm self}$=0.593.  This value although calculated at the lowest order (without any quark loops) may be compared with the value we
  get from the numerical fit to the Wilson loop data, $ a V_0 $ ranges
  between 0.63 and 0.66 for the range for $am_q$ explored.
   
The difference might be due to higher order corrections and/or
nonperturbative contributions. 

We note that in perturbation theory, the strength of the static
Coulomb potential and the static source self energy are both given by
the strong coupling constant $\alpha_s$. 
In higher order of perturbation theory \cite {bali-boyle} the static source self energy can have $am_q$-dependence, the self energy increasing with $am_q$ decreasing.  As pointed out earlier and as shown in Fig. \ref{av0}, $aV_0$ has a weak dependence on $am_q$ with the trend suggested by Ref. \cite{bali-boyle}, but, however,  at our smallest quark masses, it approximately saturates.   
\subsection{Extraction of the strong coupling constant from the plaquette}
According to Lepage and Mackenzie \cite{lm}, the strong coupling
constant $\alpha_{\rm v}(q)$ at momentum scale $q=3.41/a$ is defined via
\be
V(q)&=& -C_F ~ 4 \pi ~ \frac{\alpha_0}{q^2} ~ \left[ 1 + \alpha_0~ 
\left(  \frac{\left( 11 - \frac{2}{3}n_f\right)}{4 \pi} {\rm ln}\frac{\pi^2}{a^2
  q^2} ~ + ~ 
4 \pi d \right)\right]\nonumber \\
&=& -C_F ~ 4 \pi ~ \frac{\alpha_{\rm v}\left(q\right)}{q^2}~
\ee
where $ d = d_g + d_f = 0.37428 - 2. \times 0.00426 = 0.36576$, $C_F=4/3$ and $n_f$ is the number of flavors.

Solving for the bare coupling $\alpha_0 = g^2/4\pi$ from 
\be
\alpha_{\rm v}(q) = \alpha_0(1~+~ \alpha_0 ~C)~ ~~ {\rm with} ~~~
C=\frac{\left( 11 - \frac{2}{3}n_f \right)}{4 \pi} {\rm ln}\frac{\pi^2}{a^2
  q^2} ~ + ~ 
4 \pi d 
\ee
we get
\be
\alpha_0 = \alpha_{\rm v}(1 - C~ \alpha_{\rm v})  
\ee

From perturbation theory \cite{spitz}, for the logarithm of the
average Plaqutte ($\Box_{\rm av}$) one gets 
\be
-{\rm ln~ \Box_{\rm av}}~=~ c_1 g^2 + \left(c_2+\frac{1}{2}c_1^2\right)~g^4 
\ee
with $ c_1 =1.3$ and $ c_2 = c_2^g + c_2^f = 0.03391 - .003696 =
0.0302$. 
Thus 
\be
-{\rm ln~ \Box_{\rm av}} &=& \frac{1}{3} 4 \pi \alpha_0 + 0.0858 \times
  16 \pi^2 \alpha_0^2 \nonumber \\
& = & \frac{1}{3} 4 \pi \alpha_{\rm v} \left[ 1 - \alpha_{\rm v} \left( C - 0.0858
    \times 12 \pi \right) \right].
\ee 
\begin{figure}
\begin{center}
\includegraphics[width=.8\textwidth]{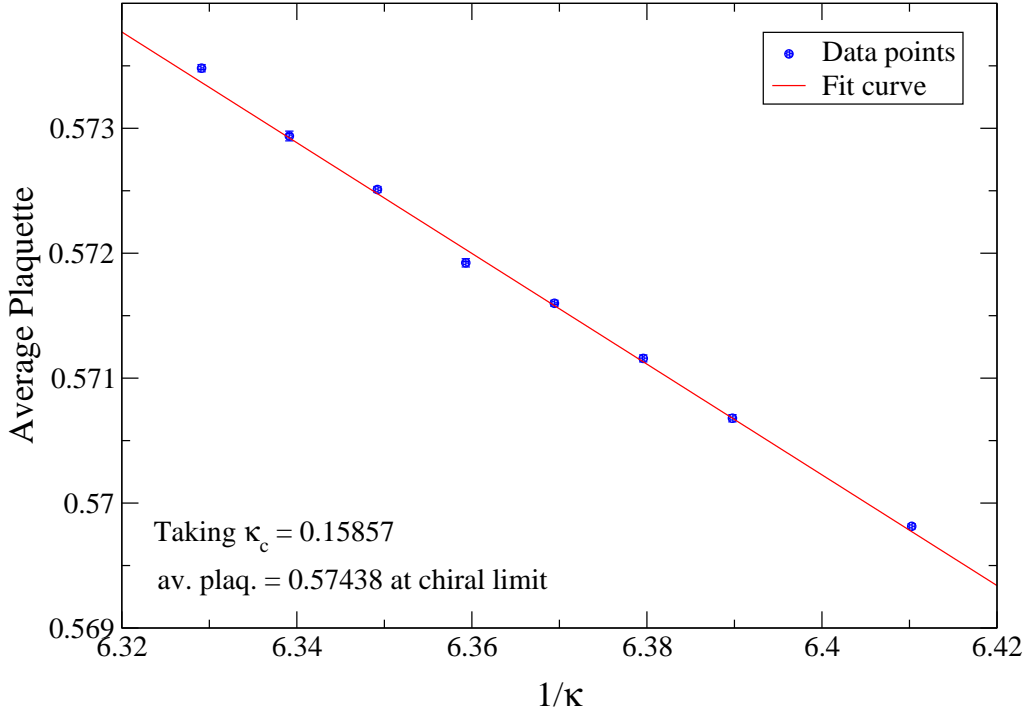}
\end{center}
\caption{Average value of the plaquette versus $1/\kappa$. }
\label{plaqfit}
\end{figure}
From the measurement of the average plaquette (${\Box_{\rm av}}$) at each $\kappa$ and extrapolating to $\kappa_c$ one 
can determine the
strong coupling constant at a given scale. As shown in Fig. \ref{plaqfit},  omitting the data points
at $\kappa$=0.156 and 0.158 we perform a linear fit of the average plaquette and find ${
  \Box_{\rm av}}$= 0.57438 at $\kappa_c$= 0.15857. Using this value,
we find $ \alpha_{\rm v} (3.41/a) = 0.167.$ If we run down the coupling to
the scale $1/a$ using two loop renormalization group formula, we get 
$\alpha_{\rm v}(1/a)=0.257$. In the same convention, the coefficient of the
$1/R$ potential from the fit (see Fig. \ref{alpha}) ${\overline \alpha}= \frac{3}{4} \times
0.30 = 0.225$.  
\section{Conclusions}\label{concl}
Understanding the dependence of the static potential and the quantities derived from it, in particular $a/r_c$, on the sea quark mass is mandatory to set the lattice scale using the potential and is also important for chiral extrapolation of hadronic observables. At present there is very little theoretical understanding of the sea quark mass dependence of the parameters $\sigma^{1/2}$ and $1/r_c$. Accurate numerical evidence at small $am_q$ of the observed dependence may be useful for theoretical understanding.

Interestingly this dependence has been observed by {\em all} lattice QCD simulations even including the improved actions like overlap and domain wall fermions along with improved gauge actions. As a result it is difficult to consider this as a cut-off effect purely.

Our approach has been to investigate the issue with an action with the most lattice artifacts, i.e., the standard Wilson fermion and gauge action, but with (i) a large enough $\beta$ (=5.6) so that the scale $a$ is small enough, and (ii) a large set (eight values) of the fermionic hopping parameter $\kappa$ for sea quarks corresponding to PCAC quark masses in lattice units $am_q$ from around 0.07 to less than 0.015. This allows us to look for lattice artifacts or scale-violations quantitatively in terms of a reasonably large range of $am_q$.

What we have found {\em numerically} is that for small $am_q$ corresponding to $am_q\lesssim 0.035 $, the quantities related to the static potential have specific orderly behavior. With the usual Cornell potential parameterization, we find that for $am_q\lesssim 0.035$, the parameter $\alpha$ (coefficient of the $1/R$ term) is independent of $am_q$ while $a\sigma^{1/2}$ (coefficient of the linear $R$ term) depends on $am_q$ linearly with a positive intercept at $am_q=0$, resulting in a linear $am_q$ dependence of the quantities $a_/r_c$ where $r_c$ is a Sommer-type scale parameter. We have taken utmost care in all aspects of the analysis to come to the above behaviors of quantities numerically, e.g., with regard to optimum smearing level for the gauge configurations at each $\kappa$, correction for $1/R$ on the finite lattice, fitting range $[T_{\rm min}, T_{\rm max}]$  of Wilson loop data, fitting range $[R_{\rm min}, R_{\rm max}]$ of the static potential data etc and we believe that these conclusions are independent of choice of parameters of the analysis at least qualitatively and to our precision even quantitatively.

With the above, now if we accept a mass-independent scheme, the above linear $am_q$ dependence of $a\sigma^{1/2}$ and $a/r_c$ naturally translates into a physical linear $m_q$ dependence of $\sigma^{1/2}$ and $1/r_c$. These dimensionful quantities are then very similar to $m_\rho$ as far as $m_q$ dependence is concerned.

Once $m_q$ dependence is taken as a physical effect, in the mass independent scheme there is then conceptually no problem taking a chiral extrapolation of the numerical data of $a\sigma^{1/2}$ and $a/r_c$ to the physical point for a dimensionful value of $\sigma^{1/2}$ and the scale $a$. 

We have exercised care also in the chiral extrapolation. The quark mass in lattice unit, viz., $am_q$, is the best quantity for a chiral extrapolation, but it is not suitable for an extrapolation to the physical point. We have first used extrapolation with respect to $(am_\pi)^2$ to make sure that we get the same limits at the chiral point, i.e., $am_q=0$ or $(am_\pi)^2=0$. This was easily achieved once large $(am_\pi)^2$ points and also (with knowledge of finite size effect from our previous work \cite{paper0}) the smallest one or two $(am_\pi)^2$ point(s) were omitted from the fits. We stress that we stick to fits only with linear power of $(am_\pi)^2$ and we prefer $(am_\pi)^2$ to $(r_cm_\pi)^2$ or to $(m_\pi/m_\rho)^2$, because (i) $r_c$ has its own quark mass dependence and (ii) $(m_\pi/m_\rho)^2$  is only approximately linear in $am_q$ even for small quark masses. 

We obtain an accurate determination of the scale by solving a quadratic equation in the scale $a$ (in fm) resulting from the linear dependence of $a/r_c$ on $(am_\pi)^2$ and putting in the values of $r_c^{\rm ph}$ and $m_\pi^{\rm ph}$ at the physical point. We quote with $\sim 1\%$ error:
\be
a\,=\,0.08041(12)(77) ~{\rm fm},~~ ~~~ a^{-1}=2.454(4)(23) ~{\rm GeV}. 
\ee
The first and the second errors are respectively the systematic and the statistical errors both of which are estimated conservatively. Out of the four jackknife statistical errors shown in the first data row of Table \ref{afit}, we quote in the above only the largest error. Also the systematic error is estimated by halving the systematic difference between the scale determinations from $a/r_0$ and $a/r_1$. 

From $(m_\pi/m_\rho)^2$ extrapolations also, we have determined the scale which is consistent with our accurate determination above but has large ($>2\%$) errors.

In order to have an independent check on the scale, we have determined the scale also from a linear $am_\rho - (am_\pi)^2$ extrapolation, a method which is fully independent of the static potential and the quantities derived from it. It is very satisfying to find the scale obtained this way comes within $1\%$ of our accuarate determination achieved with the extrapolation of $a/r_c$. Errors are, however, large ($\sim 2\%$) in this case. 

In our determination of the lattice scale from the static potential, we have assumed $r_0= 0.49$ fm. There have been a few determinations of $r_0$ from low level energy splittings of the heavy-onium systems and these values are a few percent lower than the value we have used. There have also been suggestions in the literature about the uncertainty of the value of $r_0$ because it is not a quantity directly measurable from experiments. We like to mention that the hadronic masses also have some uncertainty in their values and like $r_0$ they are also likely to change with the number of flavors. Our approach in this paper has been that given the uncertainty in the value of $r_0$, we wanted to reduce the uncertainty in the rest of the determination. In addition to the scale determination,  the quark mass dependence of $a/r_c$ is also a physics issue that one needs to understand. Anyway, our scale determination independent of the potential shows that the scale obtained is extremely close to the value obtained from the potential. This indicates that the value of $r_0$ (= 0.49 fm) used in our analysis may not suffer from major uncertainties. However, we should keep in mind that our analysis is done with 2 flavors of sea quarks.

\acknowledgments

Numerical calculations are carried out on a Cray XD1 (120 AMD
Opteron@2.2GHz) supported by the
10$^{th}$ and 11$^{th}$ Five Year Plan Projects of the Theory
Division, SINP under the
DAE, Govt. of India. This work was in part based on the MILC collaboration's
public lattice gauge theory code.
See \url{http://physics.utah.edu/~dtar/milc.html}~.



\end{document}